%% file: root.tex
\documentclass[12pt]{article}

\input{usepkg.tex}


\input{macro}

\input{title.tex}

\begin{document}

\maketitle

\input{abs}


  
  \input{intro.tex}

        \input{framework.tex}

        \input{motivations-ctrl.tex}

        \input{contributions.tex}

  \input{basic.tex}

        \input{predicates.tex}\input{lts.tex}\input{lts-ctr.tex}
\input{dtlhs-new.tex}

        \input{dtlhs-ctr.tex}

        \input{dtlhs-qcp.tex}
        \input{control-abstraction.tex}



  \input{parallel-algo.tex}

  \input{expres.tex}

        \input{related-work.tex}
  \input{conclu.tex}


  \bibliographystyle{splncs}
  \bibliography{modelchecking}

\end{document}

%% file: usepkg.tex
\usepackage{etex} 
\usepackage{datetime}

\usepackage{balance}
\usepackage{epsfig,epic,eepic}
\usepackage{graphicx}
\usepackage{multicol}
\usepackage[all]{xy}
\usepackage{multirow}
\usepackage{rotating}
\usepackage{eurosym}
\usepackage{datetime}
\usepackage{hyperref}
\usepackage{import}
\usepackage{xfrac}

\usepackage{color}
\usepackage{times}
\usepackage{eurosym}
\usepackage{longtable}
\usepackage[table]{xcolor}

\usepackage{algorithm}
\usepackage[noend]{algorithmic}
\usepackage{xspace}

\usepackage{afterpage, longtable, footnote, framed, latexsym} 

\usepackage{amsfonts}
\usepackage{amssymb}
\usepackage{amsmath}
\usepackage{pgf}
\usepackage{tikz}
\usetikzlibrary{arrows}
\usetikzlibrary{decorations.pathreplacing}
\usetikzlibrary{automata}
\usetikzlibrary{positioning}

%% file: macro.tex
\newtheorem{definition}{Definition}
\newtheorem{example}{Example}

\algsetup{linenodelimiter=.} 
\algsetup{indent=1em}

\definecolor{light-gray}{gray}{0.90}

\newcounter{theorem-backup}

\newcommand{\N}{\mathbb{N}}
\newcommand{\Z}{\mathbb{Z}}
\newcommand{\R}{\mathbb{R}}

\newcommand{\B}{\mathbb{B}}

\usepackage{listings}
\lstdefinelanguage{PseudoC}[ISO]{C++} { 
morekeywords={foreach, and, not, or, is, FIFO_Queue, HashTable, FILE, Cache}, 
morekeywords=[2]{HashInsert, Enqueue, Dequeue, next}, 
mathescape=true,
alsoletter={'},
deletestring=[b]'
}

\lstdefinelanguage{PseudoCWithNumbers}[ISO]{C++} { 
morekeywords={foreach, and, not, or, is, FIFO_Queue, HashTable, FILE, Cache}, 
morekeywords=[2]{HashInsert, Enqueue, Dequeue, next}, 
mathescape=true,
alsoletter={'},
deletestring=[b]'
}

\lstdefinelanguage{Murphi}[]{Pascal} { 
morekeywords={ruleset, rule, invariant, startstate, return, endif, endfor, endswitch, forall, endforall, exists, endexists}, 
mathescape=true, 
morestring=[b]",
morestring=[b]', 
morecomment=[s]{/*}{*/} ,
morecomment=[l]{--}
}

\lstdefinelanguage{PRISM}[ISO]{C++} { 
morekeywords={probabilistic, stochastic, const, rate, module, endmodule, init, P, U},
mathescape=true, 
alsoletter={'},
deletestring=[b]'
}

\lstdefinelanguage{Yacc}[ISO]{C++} { 
morekeywords={token, left}, 
mathescape=false,
alsoletter={'},
deletestring=[b]'
}

\lstdefinestyle{PseudoC}{
language=PseudoC,
basicstyle=\ttfamily,
tabsize=1,
showlines=false,
emptylines=*1,
breaklines=true,
breakindent=5pt,
keywordstyle=\rmfamily\bfseries,
keywordstyle=[2]\rmfamily,
commentstyle=\itshape, 
columns=fixed,
showspaces=false, 
showstringspaces=false, 
showtabs=false, 
escapechar=\%
}

\lstdefinestyle{PseudoCWithNumbers}{
language=PseudoC,
basicstyle=\ttfamily,
tabsize=1,
showlines=false,
emptylines=*1,
breaklines=true,
breakindent=5pt,
keywordstyle=\rmfamily\bfseries,
keywordstyle=[2]\rmfamily,
commentstyle=\itshape, 
columns=fixed,
showspaces=false, 
showstringspaces=false, 
showtabs=false, 
escapechar=\%,
numbersep=5pt,framexleftmargin=15pt,numbers=left,
}

\lstdefinestyle{Murphi}{
language=Murphi,
basicstyle=\ttfamily,
tabsize=1,
showlines=false,
emptylines=*1,
breaklines=true,
breakindent=5pt,
keywordstyle=\rmfamily\bfseries,
commentstyle=\itshape, 
columns=fixed,
showspaces=false, 
showstringspaces=false, 
showtabs=false,
escapechar=\%}

\lstdefinestyle{PRISM}{
language=PRISM,
basicstyle=\ttfamily,
tabsize=1,
showlines=false,
emptylines=*1,
breaklines=true,
breakindent=5pt,
keywordstyle=\rmfamily\bfseries,
commentstyle=\itshape, 
columns=fixed,
showspaces=false, 
showstringspaces=false, 
showtabs=false,
escapechar=\%}

\lstdefinestyle{Yacc}{
language=PseudoC,
basicstyle=\ttfamily,
tabsize=1,
showlines=false,
emptylines=*1,
breaklines=true,
breakindent=5pt,
keywordstyle=\rmfamily\bfseries,
keywordstyle=[2]\rmfamily,
commentstyle=\itshape, 
columns=fixed,
showspaces=false, 
showstringspaces=false, 
showtabs=false, 
}

\newcommand{\qks}{\mbox{\sl QKS}}

\newcommand{\pqks}{\mbox{\sl PQKS}}

\newcommand{\fun}[1]{{\textsl{#1}}}

\definecolor{Blue}{rgb}{0.25,0.33,0.77}
\definecolor{Red}{rgb}{1,0,0}
\definecolor{Green}{rgb}{0,1,0}

\definecolor{light-gray}{gray}{0.90}



%% file: title.tex
\title{A Map-Reduce Parallel Approach to Automatic Synthesis of Control Software}

\author{Vadim Alimghuzin\inst{1} \inst{2}
  \and Federico Mari\inst{1}
  \and Igor Melatti\inst{1}
  \and Ivano Salvo\inst{1}
  \and Enrico Tronci\inst{1}
}

\author{Vadim Alimghuzin, Federico Mari, Igor Melatti, Ivano Salvo, \\Enrico Tronci\\
\small \itshape Department of Computer Science\\
\small \itshape Sapienza University of Rome\\
\small \itshape via Salaria 113, 00198 Rome\\
\small email: \{alimghuzin,mari,melatti,salvo,tronci\}@di.uniroma1.it\\
\small \itshape Vadim Alimghuzin is also with the \\
\small \itshape Department of Computer Science and Robotics \\
\small \itshape Ufa State Aviation Technical University \\
\small \itshape Russian Federation\\
Submitted to SPIN 2013
}

%% file: abs.tex
\begin{abstract}

Many Control Systems are indeed Software Based Control Systems, i.e. control
systems whose controller consists of control software running on a
microcontroller device. This motivates investigation on Formal Model Based
Design approaches for automatic synthesis of control software. 

Available algorithms and tools (e.g., \qks) may require weeks or even months of
computation to synthesize control software for large-size systems.  This
motivates search for parallel algorithms  for control software synthesis. 

\sloppy

In this paper, we present a Map-Reduce style parallel algorithm for control
software synthesis when the controlled system ({\em plant}) is modeled as
a discrete time linear hybrid system. Furthermore we present an MPI-based
implementation \pqks\ of our algorithm. To the best of our knowledge, this is
the first parallel approach for control software synthesis. 

\fussy

We experimentally show effectiveness of \pqks\ on two classical control
synthesis problems: the inverted pendulum and the multi-input buck DC/DC
converter. Experiments show that \pqks\ efficiency is above 65\%. As an example,
\pqks\ requires about 16 hours to complete the  synthesis of control software
for the pendulum on a cluster with 60 processors, instead of the 25 days needed
by the sequential algorithm implemented in \qks.

\end{abstract}

%% file: intro.tex
\section{Introduction}
%

%% file: framework.tex


Many Embedded Systems are indeed Software Based Control Systems (SBCSs). An SBCS
consists of two main subsystems: the controller and the plant. Typically, the
plant is a physical system consisting, for example, of mechanical or electrical
devices whereas the controller consists of control software running on a
microcontroller. 
In an endless loop, at discrete time instants ({\em sampling}), the
controller reads
plant sensor outputs from the plant and computes
commands to be sent back to plant actuators. 
Being the control software discrete and the physical system typically continuous, 
sensor outputs go through an Analog-to-Digital (AD) conversion ({\em quantization}) before
being read from the control software. 
Analogously,
controller commands need a Digital-to-Analog (DA) conversion before being sent to plant actuators.
The controller selects commands in order to
guarantee that the closed-loop system (that is, the system consisting of both
plant and controller) meets given safety and liveness specifications (System
Level Formal Specifications).

Software generation from models and formal specifications forms the core of
Model Based Design of embedded software \cite{Henzinger-Sifakis-fm06}. 
This approach is particularly
interesting for SBCSs since in such a case system level (formal) specifications
are much easier to define than the control software behavior itself.

%% file: motivations-ctrl.tex
\subsection{Motivations}\label{motivations.subsec}

In this paper we focus on the algorithm presented in~\cite{cav10,tosem13,cav2010-tekrep-art-2011}, 
which returns correct-by-construction control software starting from system level formal specifications.
This algorithm is implemented in \qks\ ({\em Quantized Kontroller Synthesizer}), which takes as input:
i) a formal model of the
controlled system, modeled as a Discrete Time Linear Hybrid System (DTLHS), 
ii) safety and
liveness requirements (goal region) and iii) $b$, $b_u$ as the number of bits for AD (resp., DA) conversion. Given this,
\qks\ outputs a correct-by-construction control software together with the controlled
region on which the software is guaranteed to work.

\begin{figure}
   \vspace*{-25mm}
  \includegraphics[scale=0.4]{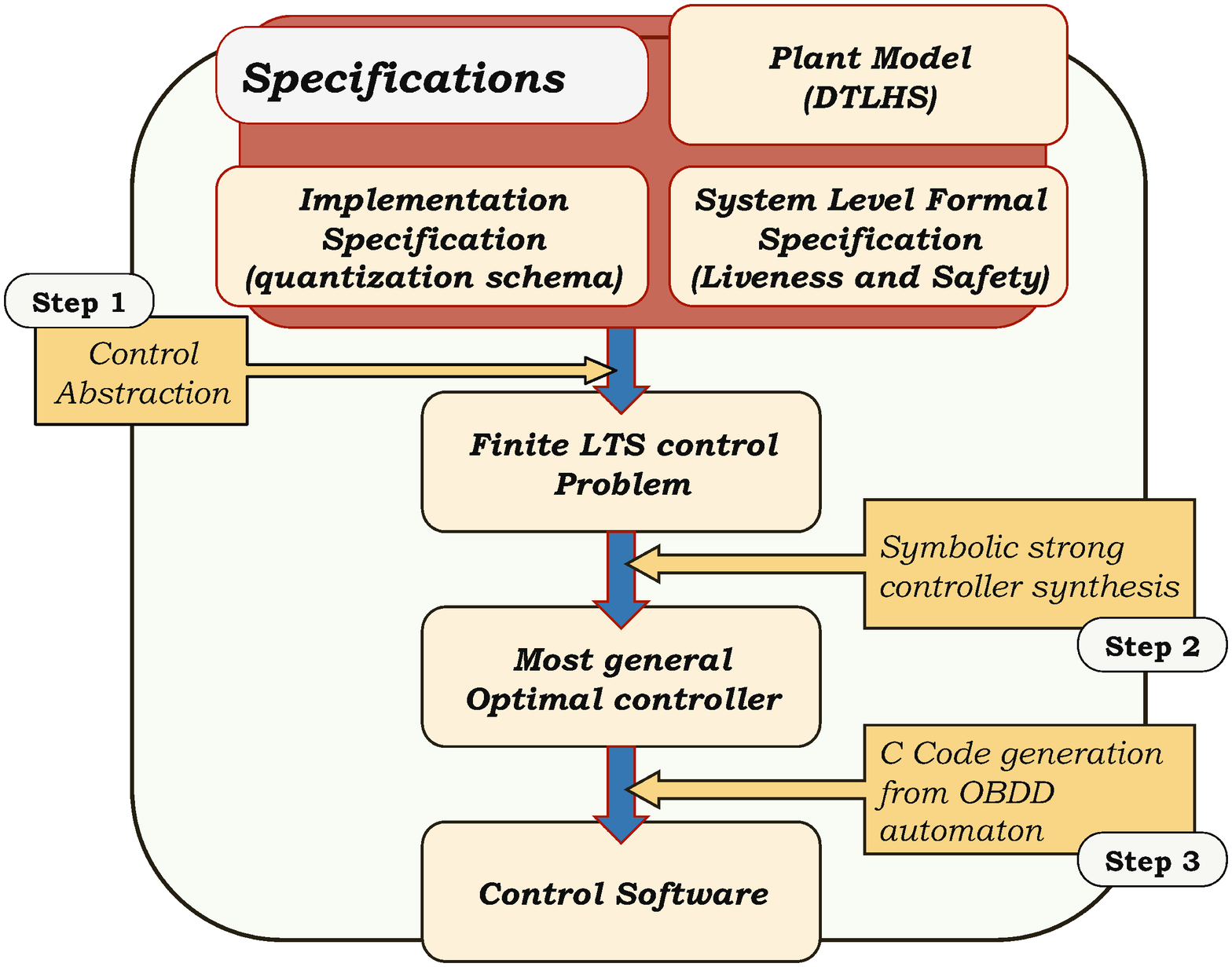}
  \caption{Control Software Synthesis Flow.}
  \label{fig:cssf}
\end{figure}

To this aim, \qks\ first computes a suitable finite state abstraction 
({\em control abstraction}~\cite{cav2010-tekrep-art-2011}) $\hat{\cal H}$ 
of the DTLHS plant model ${\cal H}$, where
$\hat{\cal H}$ depends on the quantization schema (i.e. number of bits $b$ needed for AD conversion)
and it is the plant as it can be seen from the control software after AD conversion.
Then, given an abstraction $\hat{G}$ of the goal states $G$, it is computed 
a controller $\hat{K}$ that, starting from any initial abstract state,
 drives $\hat{\cal H}$ to $\hat{G}$ regardless of possible nondeterminism.
Control abstraction properties ensure that $\hat{K}$ 
is indeed a (quantized representation of a) controller for the original plant ${\cal H}$.
Finally, the finite state automaton $\hat{K}$ is translated into control software (C code).
The whole process is depicted in Fig.~\ref{fig:cssf}. 

While effective on moderate-size systems, \qks\ requires a huge amount of computational resources 
when applied to larger systems. 
In fact, 
the most critical step of \qks\ is
the control abstraction $\hat{\cal H}$ generation (which is responsible for more
than 95\% of the overall computation, see~\cite{tosem13}).  This stems from the fact that
$\hat{\cal H}$  is computed explicitly, by solving a Mixed Integer Linear Programming (MILP) problem for each triple
$(\hat{x}, \hat{u}, \hat{x}')$, where 
$\hat{x}, \hat{x}'$ are abstract states of $\hat{\cal H}$  and $\hat{u}$ is an abstract
action of $\hat{\cal H}$.  Since the number of abstract states is $2^b$, being
$b$ the number of bits needed for AD conversion of all variables describing the plant,
we have that \qks\  computation time is exponential in
$2b + b_u$. 
In \qks, suitable optimizations reduce the complexity to be exponential in $b +
b_u$, and thus in $b$ since $b_u << b$. However, 
in large-size systems 
$b$ may be large for two typical reasons. 
First, since each plant
state variable needs to be quantized (if a state variable $v$ is discrete, then the number of bits for $v$
is not an input, since 
$\lfloor\log_2 |{\rm dom}(v)|\rfloor + 1$ bits are needed), the number of bits is
necessarily high when the plant model consists of many variables.
As an example, the plane collision avoidance 
control system in~\cite{TLS99} is described by 4 continuous variables and 7 discrete variables. 
Second, controllers synthesized by considering a 
finer quantization schema (i.e., with an higher value of $b$) usually have a better behavior with respect to 
non-functional requirements, such as {\em ripple} and 
{\em set-up time}. 
Therefore, when a high precision is required, 
a large number of quantization bits 
must be considered. 

As an example, experimental
results show that \qks\ takes nearly one month (25 days) of CPU time
to synthesize the controller 
for a 26 bits quantized
inverted pendulum (which is described by only two continuous state variables, see Sect.~\ref{sect:invpend}). 
Moreover, 99\% of those 25 days of computation is due to control abstraction
generation. 
This may result in a loss in terms of time-to-market in 
control software design when \qks\ is used.

This motivates search of parallel versions of \qks\  synthesis algorithm.


%% file: contributions.tex
\subsection{Main Contributions}

To overcome the computation time bottleneck in \qks, 
we present a {\em Map-Reduce} style parallel algorithm for control abstraction generation 
in control software synthesis.

Map-Reduce~\cite{map-red08} is a (LISP inspired) programming paradigm advocating a form of embarrassing parallelism
for effective massive parallel processing. An implementation of such an approach is in Hadoop (e.g., see~\cite{mapreduce.book}).
The effectiveness of the Map-Reduce approach stems from the minimal communication overhead of embarrassing parallelism.
This motivates our goal of looking for a map-reduce style parallel algorithm for control software synthesis from system level
formal specifications.

To this aim, we design a parallel version of \qks, that is inspired to the
Map-Reduce programming style and that we call {\em Parallel
\qks} (\pqks\ in the following). 
\pqks\ is actually implemented using MPI
(Message Passing Interface~\cite{mpi.book}) 
in order to exploit the computational power available in modern computer
clusters (distributed memory model).
Such an algorithm will be presented in
Sect.~\ref{parallel-algo.tex}, after a discussion of the basic notions needed
to understand our approach (Sect.~\ref{basic.tex}) and the description of the
standalone (i.e. serial) algorithm of \qks\ (Sect.~\ref{sec:control-abstraction}).

We show the effectiveness of \pqks\ by using it 
to synthesize control software for two widely used embedded systems, namely
the multi-input buck DC-DC converter~\cite{multin-buck-dcdc-2010} and the
inverted pendulum~\cite{KB94} benchmarks. 
These are challenging examples for the automatic synthesis of correct-by-construction 
control software.
Experimental results on the
above described benchmarks will be
discussed in Sect.~\ref{expres.tex}. Such results show that we achieve a nearly linear speedup
w.r.t. \qks, with efficiency above 65\%. 
As an example, \pqks\ 
requires about 16 hours to complete the above mentioned 
synthesis of the 26-bits pendulum on a cluster with 60 processors,
instead of the 25 days of \qks.

%
%

%% file: basic.tex
%
%
%
%
%

%
%
\section{Background on DTLHS Control Software Synthesis} \label{basic.tex}
To make this paper self-contained, 
in this section we briefly summarize previous work on automatic generation of 
control software for {\em Discrete Time Linear Hybrid System} (DTLHS) 
from System Level Formal Specifications. 

As shown in \figurename~\ref{fig:cssf}, 
we model the controlled system (i.e. the plant) as a DTLHS (Sect.~\ref{sec:dtlhs}), 
that is a discrete time hybrid system 
whose dynamics is modeled as a 
{\em guarded (linear) predicate} (Sect.~\ref{subsection:predicates})  
over a set of continuous as well as discrete 
variables.
The semantics of a DTLHS is given in terms of a {\em Labeled Transition Systems} 
(LTS, Sect.~\ref{sec:lts}).
Given a DTLHS plant model ${\cal H}$, a set of {\em goal states} $G$ 
({\em liveness specifications}) and an {\em initial region} $I$, 
both represented as linear predicates, 
we are interested in finding a {\em restriction $K$ of the behaviour} 
of ${\cal H}$ such that in the {\em closed loop system} 
all paths starting in a state in $I$ lead to $G$ after a finite number of steps. 
Finding $K$ is the DTLHS {\em control problem} (Sect.~\ref{sec:dtlhs-ctr}) 
that is in 
turn defined as a 
suitable LTS control problem.  (Sect.~\ref{sec:lts-ctr}).
Since we want to output a control software, we are interested in controllers that take their decisions by looking at 
{\em quantized states}, i.e. the values that the control software 
reads after an AD conversion. To this aim, the solution of a {\em quantized control problem} 
(Sect.~\ref{sec:qcp}) is computed by first generating a discrete abstraction of ${\cal H}$, 
called {\em control abstraction} (Sect.~\ref{sec:control-abstraction}, step 1
in \figurename~\ref{fig:cssf}), then by applying to such control abstraction known
techniques in order to generate a controller (step 2 in
\figurename~\ref{fig:cssf}), and finally synthesizing a control software (step 3 in
\figurename~\ref{fig:cssf}). Our main contribution in this paper is in the control
abstraction generation, thus we will focus this section on the basic notions to
understand definition and
computation of control
abstractions (Sect.~\ref{sec:control-abstraction}).

%% file: predicates.tex
\subsection{Predicates}
\label{subsection:predicates}
We denote with $[n]$ an initial segment $\{1,\ldots, n\}$ of the natural numbers. 
We denote with $X$ = $[x_1, \ldots, x_n]$ a
finite sequence of variables that we may regard, when convenient, as a set.
Each variable $x$ ranges on a known (bounded or unbounded)
interval ${\cal D}_x$ either of the reals (continuous variables) 
or of the integers (discrete
variables). 
We denote with ${\cal D}_X$ the set $\prod_{x\in X} {\cal D}_x$.
Boolean variables are discrete variables
ranging on the set $\B$ = \{0, 1\}. 
To clarify that a variable $x$ is {\em continuous} 
(resp. {\em discrete}, resp. {\em boolean}) we may write $x^{r}$ (resp. $x^d$, $x^b$). 
Analogously $X^{r}$ ($X^{d}$, $X^{b}$) denotes the sequence
of real (discrete, boolean) variables in $X$.
Unless otherwise stated, we suppose
${\cal D}_{X^r} = \R^{|X^r|}$ and ${\cal D}_{X^d} = \Z^{|X^d|}$. 
If $x$ is a boolean variable,  
we write $\bar{x}$ for $(1 - x)$.

A {\em linear expression} over a list of variables $X$ is a linear
combination of variables in $X$ with rational coefficients. 
A {\em linear constraint} over $X$ (or simply a {\em constraint})  is an expression of the form
$L(X) \leq b$,
where $L(X)$ is a linear expression over $X$ 
and $b$ is a rational constant. In the following, we also write $L(X) \geq b$
for $-L(X) \leq -b$.

{\em Predicates} are inductively defined as follows.
A constraint $C(X)$ over a list of variables $X$ is a predicate over 
$X$. 
If $A(X)$ and $B(X)$ are predicates over $X$, then $(A(X) \land B(X))$
and $(A(X) \lor B(X))$ are predicates over X.  Parentheses may be
omitted, assuming usual associativity and precedence rules of logical
operators.
A {\em conjunctive predicate} is a conjunction of constraints.
For conjunctive predicates we will also write: 
$L(X) = b$ for (($L(X) \leq b$) $\wedge$ ($L(X) \geq b$)) and $a \leq x \leq
b$ for $x \geq a \;\land\; x \leq
b$, where $x \in X$.

Given a constraint $C(X)$ and a fresh boolean variable ({\em guard}) $y \not\in X$,
the {\em guarded constraint} $y \to C(X)$ (if $y$ then $C(X)$) denotes
the predicate $((y = 0) \lor C(X))$. Similarly, we use $\bar{y} \to
C(X)$ (if not $y$ then $C(X)$) to denote the predicate $((y = 1) \lor
C(X))$.
A {\em guarded predicate} is a conjunction of 
either constraints or guarded constraints.

%% file: lts.tex
\subsection{Labeled Transition Systems}
\label{lts.tex}
\label{sec:lts}

A \emph{Labeled Transition System} (LTS) is a tuple
${\cal S} = (S, {\cal A}, T)$ where 
$S$ is a (possibly infinite) set of states, 
${\cal A}$ is a (possibly infinite) set of \emph{actions}, and 
$T$ : $S$ $\times$ ${\cal A}$ $\times$ $S$ $\rightarrow$ $\B$
is the \emph{transition relation} of ${\cal S}$.
We say that $T$ (and ${\cal S}$) is {\em deterministic} if $T(s, a, s') \land
T(s, a, s'')$ implies $s' = s''$, and {\em nondeterministic}
otherwise. 
Let $s \in S$ and $a \in {\cal A}$.
We denote with 
$\mbox{\rm Adm}({\cal S}, s)$ the set of actions
admissible in $s$, that is $\mbox{\rm Adm}({\cal S}, s)$ = $\{a \in {\cal A}
\; | \; \exists s': T(s, a, s') \}$
and with
$\mbox{\rm Img}({\cal S}, s, a)$ the set of next
states from $s$ via $a$, that is $\mbox{\rm Img}({\cal S}, s, a)$ =
$\{s' \in S \; | \; T(s, a, s') \}$.
We call {\em self-loop} a transition of the form $T(s, a, s)$.
A {\em run} or \emph{path}
for an LTS ${\cal S}$ 
is a sequence 
$\pi$ =
$s_0, a_0, s_1, a_1, s_2, a_2, \ldots$ 
of states $s_t$ and actions $a_t$ 
such that
$\forall t \geq 0$ $T(s_t, a_t, s_{t+1})$.
The length $|\pi|$ of a finite run $\pi$ is the number of actions
in $\pi$. 
Sometimes $s_t$ (resp. $a_t$) will be denoted by $\pi^{(S)}(t)$ (resp. $\pi^{({\cal A})}(t)$). 

%% file: lts-ctr.tex
\subsection{LTS Control Problem and Solutions}
\label{sec:lts-ctr}
A \emph{controller} for an LTS ${\cal S}$ 
is used to restrict the dynamics of ${\cal S}$ 
so that all states in the initial region 
will reach 
the goal region. 
In the following, we formalize such a concept
by defining 
solutions to an LTS control problem. 
In what follows, let ${\cal S} = (S, {\cal A}, T)$ be an LTS, 
$I$, $G$ $\subseteq$ $S$ 
be, respectively, the {\em initial} and {\em goal} regions of ${\cal S}$.

\begin{definition}
  \label{def:ctroller-lts}
  \label{def:ctrproblem-lts}
  A \emph{controller} for 
${\cal S}$ is a function 
  $K\! :\! S \times {\cal A}\! \rightarrow\! \B$
  such that $\forall s \in S$, $ \forall a \in {\cal A}$, if $K(s, a)$ then
  $\exists s' \; T(s, a, s')$. 
  If $K(s,a)$ holds, we say that the action $a$ is {\em enabled} by $K$ in $s$. 

 The set of states $\{s \in
  S \; | \; $$\exists a \; K(s, a)\}$ for which at least an 
  action is enabled is denoted by ${\rm dom}(K)$.

${\cal S}^{(K)}$ denotes the \emph{closed loop system}, that
  is the LTS $(S, {\cal A}, T^{(K)})$, where 
$T^{(K)}(s,$ $a,$ $s')$ $=$ $T(s, a, s') \wedge K(s, a)$.
\end{definition}
\smallskip

We call a path $\pi$ {\em fullpath} 
 if either it is infinite or its last state 
$\pi^{(S)}(|\pi|)$ has no successors 
(i.e. $\mbox{\rm Adm}({\cal S}, \pi^{(S)}(|\pi|)) = \varnothing$).
${\rm Path}(s, a)$ denotes the set of fullpaths starting in state
$s$ with action $a$, i.e. the set of fullpaths $\pi$ s.t. $\pi^{(S)}(0)=s$
and $\pi^{({\cal A})}(0)=a$.
Given a path $\pi$ in ${\cal S}$, 
we define $j({\cal S},\pi,G)$ as follows. If there exists $n > 0$ s.t.
$\pi^{(S)}(n)\in G$, then $j({\cal S},\pi,G) = \min\{n \;|\; n >
0 \land \pi^{(S)}(n)\in G\}$. Otherwise, $j({\cal S},\pi,G) = +\infty$.
We require $n > 0$ since
our systems are nonterminating and each controllable state (including a goal state)
must have a path of positive length to a goal state.
Taking ${\rm sup}\, \varnothing = +\infty$, 
the {\em worst case distance} 
of a state $s$ from the goal region  $G$
is  $J({\cal S},G,s)={\rm sup} \{j({\cal S},G,\pi)~|~  \pi \in{\rm Path}(s,a), a \in {\rm
Adm}({\cal S},s)\}$.

\input{lts1-pic.tex}
\input{lts2-pic.tex}

\begin{definition}
\label{def:sol}
An LTS \emph{control problem} 
is a triple 
  $\cal P$ = $({\cal S},$ $I,$ $G)$. 
A {\em strong 
solution} (or simply a solution) to 
${\cal P}$ 
is 
a controller $K$ for ${\cal S}$, such that $I$ $\subseteq$ ${\rm dom}(K)$ and
for all $s \in {\rm dom}(K)$, 
$J({\cal S}^{(K)}, G, s)$ 
is finite.

A solution 
$K^{*}$ to ${\cal P}$ is \emph{optimal} if for all solutions
$K$ to ${\cal P}$, 
for all $s \in S$, we have 
$J({\cal S}^{(K^{*})}, G, s) \leq J({\cal S}^{(K)}, G, s)$.
\end{definition}
\smallskip

\begin{example}
\label{ex:solution}
Let ${\cal S}_1=(S_1,{\cal A}_1,T_1)$ be the LTS in Fig.~\ref{fig:lts-ex-1} 
and let ${\cal S}_2=(S_2,{\cal A}_2,T_2)$ be the LTS in Fig.~\ref{fig:lts-ex-2}.
$S_1$ is the integer interval $[-1,2]$ and $S_2=[-2,5]$. 
${\cal A}_1={\cal A}_2=\{0,1\}$ and the transition relations $T_1$ and $T_2$ are  
defined by all solid arrows in the pictures. 
Let $I_1=S_1$, $I_2=S_2$ and let $G=\{ 0 \}$. 
There is no solution to the control problem $({\cal S}_1, I_1, G)$. 
Because of the self-loops of the state 1, we have that both $j({\cal S}_1, G, 1,0)$ $=$ $+\infty$
and $j({\cal S}_1, G, 1,1)$ $=$ $+\infty$. 
The controller $K_2$ defined by $K_2(s,a)\equiv((s=1\,\lor\,s=2)\,\land\,a=1)\,\lor\,(s\not=1\,
\land\,s\not=2\,\land\,a=0)$ 
is an optimal strong solution for the control problem $({\cal S}_2, I_2, G)$. 
\end{example}

%% file: lts1-pic.tex
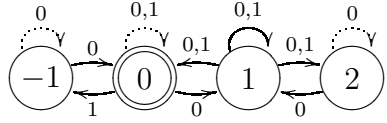
\begin{figure}
\centering
$
 \xymatrix@C=5mm@R=5mm{
%
	*+=<25pt>[o][F-]{-1} 
		\ar@(ul,ur)@{.>}[]^{0}
 		\ar@/^/[r]^{0}
    &
    *+=<25pt>[o][F=]{0} 
		\ar@(ul,ur)@{.>}[]^{0,1}
		\ar@/_/[r]_{0}		
		\ar@/^/[l]^{1}		
    &
	*+=<25pt>[o][F-]{1} 
		\ar@/_/[l]_{0,1}
		\ar@/^/[r]^{0,1}		
		\ar@(ul,ur)@[]^{0,1}
    & 
    *+=<25pt>[o][F-]{2} 
		\ar@/^/[l]^{0}
		\ar@(ul,ur)@{.>}[]^{0}
	} 
 $
 \caption{The LTS ${\cal S}_1$ in Example~\ref{ex:solution}.} 
 \label{fig:lts-ex-1}
 \end{figure}

%% file: lts2-pic.tex
\begin{figure}
\centering
$
 \xymatrix@C=5mm@R=5mm{
%
    &
    *+=<25pt>[o][F=]{0} 
		\ar@(ul,ur)@{.>}[]^{0,1}
		\ar@/_/[r]_{0}		
		\ar@/^/[d]^{1}		
    &
	*+=<25pt>[o][F-]{1} 
		\ar@/_/[l]_{1}
		\ar@/_/[r]_{0}		
		\ar@(ul,ur)@{.>}[]^{0,1}
    &
	*+=<25pt>[o][F-]{2} 
		\ar@/_/[l]_{0,1}
		\ar@/^/[r]^{0}		
		\ar@(ul,ur)@[]^{0}
		\ar@(dl,dr)@{.>}[]_{1}
    & 
    *+=<25pt>[o][F-]{3} 
		\ar@/^/[l]^{0,1}
		\ar@/^/[d]^{1}		
		\ar@(ul,ur)@[]^{1}
		\ar@(ur,dr)@{.>}[]^{0}
	\\
	& 
	*+=<25pt>[o][F-]{-1} 
		\ar@(dl,dr)@{.>}[]_{0,1}
 		\ar@/^/[u]^{0}
		\ar@/_/[r]_{1}		
	&
	*+=<25pt>[o][F-]{-2} 
		\ar@/_/[l]_{0}
		\ar@(dl,dr)@{.>}[]_{0}
    &
	*+=<25pt>[o][F-]{5} 
		\ar@/_/[r]_{0}
		\ar@(dl,dr)@{.>}[]_{0}
	& 
	*+=<25pt>[o][F-]{4} 
		\ar@/^/[u]^{0}
		\ar@/_/[l]_{1}
		\ar@(dl,dr)@{.>}[]_{0,1}
	&
	\\
	} 
 $
 \caption{The LTS ${\cal S}_2$ in Example~\ref{ex:solution}.} 
 \label{fig:lts-ex-2}
\end{figure}

%% file: dtlhs-new.tex
\subsection{Discrete Time Linear Hybrid Systems}
\label{dths.tex}
\label{sec:dtlhs}
In this section we introduce the class of 
discrete time Hybrid Systems that 
we use as plant models, namely 
{\em Discrete Time Linear Hybrid Systems} (DTLHSs for short).

\begin{definition}
\label{dths.def}

A {\em Discrete Time Linear Hybrid System} 
is a tuple ${\cal H} = (X,$ $U,$ $Y,$ $N)$ where:

\begin{itemize}
\item 
  $X$ = $X^{r} \cup X^{d}$ 
  is a finite sequence of real ($X^{r}$) and 
  discrete ($X^{d}$) 
  {\em present state} variables.  
  We denote with $X'$ the sequence of 
  {\em next state} variables obtained 
  by decorating with $'$ all variables in $X$.

\item 
  $U$ = $U^{r} \cup U^{d}$ 
  is a finite sequence of 
  \emph{input} variables.

\item 
  $Y$ = $Y^{r} \cup Y^{d}$ 
  is a finite sequence of
  \emph{auxiliary} variables that are typically used to
  model \emph{modes} (e.g., from switching elements such as diodes) 
  or ``local'' variables.

\item 
  $N(X, U, Y, X')$ is a guarded predicate 
  over $X \cup U \cup Y \cup X'$ defining the 
  {\em transition relation} (\emph{next state}). 
\end{itemize}

\end{definition}

The semantics of DTLHSs is given in terms of LTSs. 

\begin{definition}
Let ${\cal H}$ = ($X$, $U$, $Y$, $N$) 
be a DTLHS.
The dynamics of ${\cal H}$ 
is defined by the Labeled Transition System 
$\mbox{\rm LTS}({\cal H})$ = (${\cal D}_X$, ${\cal D}_U$,
$\tilde{N}$) where:
$\tilde{N} : {\cal D}_X \; \times \; {\cal D}_U \; \times \; {\cal D}_X \rightarrow \B$ 
is a function s.t.  $\tilde{N}(x, u, x') \equiv \exists \; y \in {\cal D}_Y \; N(x, u, y, x')$.
A \emph{state} $x$ for ${\cal H}$ is a state $x$ for 
$\mbox{\rm LTS}({\cal H})$ and a \emph{run} 
(or \emph{path}) for ${\cal H}$ is
a run for $\mbox{\rm LTS}({\cal H})$ (Sect. \ref{lts.tex}).
\end{definition}

%% file: dtlhs-ctr.tex
\subsection{DTLHS Control Problem}
\label{sec:dtlhs-ctr}
\vspace{-0.75mm}
A DTLHS 
control problem $({\cal H}, I, G)$ is defined as the LTS
control problem ($\mbox{\rm LTS}(\cal H)$, $I$, $G$).
To accommodate quantization errors,
always present in software based controllers, it is useful
to relax the notion of solution by tolerating
an arbitrarily small error $\varepsilon$ on the continuous variables.

Let $\varepsilon>0$ be a real number, 
$W\subseteq \R^n\times \Z^m$. 
The $\varepsilon$-\emph{relaxation} of $W$ is the \emph{ball of radius} $\varepsilon$ 
${\cal B}_{\varepsilon}(W)$
= \{($z_1, \ldots z_n$, $q_1, \ldots q_m$) 
$|$ $\exists (x_1, \ldots, x_n, q_1, \ldots q_m) \;
\in \; W$ and $\forall i \in [n] \; 
|z_i - x_i| \leq \varepsilon$\}.

\begin{definition}
  \label{def:dtlhs-ctr-prb}
  Let $({\cal H}, I, G)$ be a DTLHS control problem and $\varepsilon$
be a nonnegative real number.
%
%
%
%
%
%
%
%
%
%
%
%
%
%
An $\varepsilon$ solution to $({\cal H}, I, G)$ is a 
solution to the LTS control problem 
$(\mathrm{LTS}({\cal H}),$ $I,$ ${\cal B}_{\varepsilon}(G))$. 
\end{definition}

\begin{example}
\label{ex:ctr-dths}
Let T be the positive constant $\sfrac{1}{10}$ (sampling time).
We define the DTLHS ${\cal H}$ $=$ $(\{x\},\{u\},$ $\varnothing$, $N)$ where 
$x$ is a continuous variable, $u$ is boolean, and
$N(x, u, x')$ $\equiv$
$[\overline{u} \rightarrow x' = x+(\sfrac{5}{4}-x)T] 
\land
[u \rightarrow x' = x+ (x - \sfrac{7}{4})T]$.
Let $I(x) \equiv -1\leq x\leq \sfrac{5}{2}$ and $G(x) \equiv x=0$.
Finally, let  ${\cal P}$ be the control problem (${\cal H}$, $I$, $G$). 
A controller may drive the system near to the goal $G$, 
by enabling a suitable action in such 
a way that $x'<x$ when $x>0$ and $x'>x$ when $x<0$.
However the controller $K(x,u)$ defined by 
$
K(x,u)\equiv (-1\leq x< 0\;\land\; \overline{u}) \; \lor \; 
(0\leq x< 2 \;\land\; u) \; \lor \;
(1\leq x\leq \sfrac{5}{2} \; \land\; \overline{u})
$ 
is not a solution, because it allows infinite paths
to be executed. Since $K(\sfrac{5}{4}, 0)$ and 
$N(\sfrac{5}{4},0,\sfrac{5}{4})$ hold, 
the closed loop system ${\cal H}^{(K)}$ may loop forever along 
the path $\sfrac{5}{4},0,\sfrac{5}{4},0\ldots$.
$K'$ defined by 
$
K'(x,u)\equiv(-1\leq x< 0\;\land\; \overline{u}) \; \lor \; 
(0\leq x\leq \sfrac{3}{2} \;\land\; u) \; \lor \;
(\sfrac{3}{2}\leq x\leq \sfrac{5}{2} \; \land\; \overline{u})
$ is a solution to ${\cal P}$. 
\end{example}

%% file: dtlhs-qcp.tex
\subsection{Quantized Control Problem}
\label{sec:qcp}
\vspace{-0.75mm}
As usual 
in classical control theory, 
{\em quantization} (e.g., see
\cite{quantized-ctr-tac05}) is the process of
approximating a continuous interval by a set of integer values. 
In the following we formally define the quantized feedback control
problem for DTLHSs.

A {\em quantization function} $\gamma$
for a real interval $I=[a,b]$ is a non-decreasing   
function $\gamma:I\mapsto \Z$ s.t. $\gamma(I)$ is a bounded integer interval.
We will denote $\gamma(I)$ as $\hat{I}=[\gamma(a),\gamma(b)]$.
The \emph{quantization step} of $\gamma$, notation $\|\gamma\|$, 
is defined as ${\rm sup}\{ \; |w-z|
  \; | \; w, z \in I \land \gamma(w)=\gamma(z)\}$. 
For ease of notation, we extend quantizations to integer intervals,
by stipulating that in such a case the quantization function
is the identity function.

\begin{definition}
  Let ${\cal H}\! =\! (X, U, Y, N)$ be a DTLHS, and 
  $W=X\cup U\cup Y$.   
  A \emph{quantization} ${\cal Q}$ for $\cal H$
  is a pair $(A, \Gamma)$, where:
  \begin{itemize}

  \item
  $A$ is a 
  predicate over $W$ 
  that explicitely bounds each variable in $W$ (i.e., $A = \bigwedge_{w \in W}
  \alpha_w \leq w \leq \beta_w$, with $\alpha_w, \beta_w \in {\cal D}_W$). 
  For each $w\in W$, we denote  
  with $A_w = [\alpha_w, \beta_w]$ its {\em admissible region} and with $A_W$ $=$ $\prod_{w\in W} A_w$.
%
%
  \item 
  $\Gamma$ is a set of maps $\Gamma = \{\gamma_w$ $|$
  $w \in W$ and $\gamma_w$ is a 
  quantization function for $A_w\}$. 
  \end{itemize}
  Let $W = [w_1, \ldots w_k]$ 
  and $v = [v_1, \ldots v_k] \in A_{W}$. 
  We write $\Gamma(v)$ for the tuple $[\gamma_{w_1}(v_1), 
  \ldots \gamma_{w_k}(v_k)]$. 
  Finally, the \emph{quantization step} $\|\Gamma\|$ is 
  defined as ${\rm sup} \{ \; \|\gamma\| \; | \; \gamma \in
  \Gamma \}$.
\end{definition}
\smallskip

A control problem admits a \emph{quantized solution} if control
decisions can be made by just looking at quantized values. This enables
a software implementation for a controller. 

\begin{definition}
  \label{def:qfc}
  Let ${\cal H} = (X, U, Y, N)$ be a DTLHS,
  ${\cal Q}=(A,\Gamma)$ be a quantization for ${\cal H}$
  and ${\cal P} = ({\cal H}, I, G)$ be a DTLHS control problem.
  A ${\cal Q}$ \emph{Quantized Feedback Control} (QFC) 
  solution to ${\cal P}$ is a
  $\|\Gamma\|$ 
  solution $K(x, u)$ to ${\cal P}$ such that
  $
  K(x, u) = \hat{K}(\Gamma(x), \Gamma(u))
  $
  where 
  $\hat{K} : \Gamma(A_{X}) \times \Gamma(A_{U})$ $\rightarrow$ $\B$.
\end{definition}

\begin{example}
\label{ex:q-ctr}
Let ${\cal P}$, $K$ and $K'$ be as in Ex.~\ref{ex:ctr-dths}.
Let us consider the quantizations ${\cal Q}_1=(A_1, \Gamma_1)$, where 
$A_1=I$, 
$\Gamma_1$ = $\{\gamma_x\}$ and $\gamma_x(x)=\lfloor x\rfloor$.
The set $\Gamma(A_x)$ of quantized states is the integer interval $[-1,2]$. 
No ${\cal Q}$ QFC solution can exist, because in state $1$ either enabling action $1$ or action  
$0$ 
allows infinite loops to be potentially executed in 
the closed loop system. 
The controller $K'$ in Ex.~\ref{ex:ctr-dths} can be 
obtained as a quantized controller decreasing the quantization step,  
for example, by considering the quantization 
${\cal Q}_2=(A_2, \Gamma_2)$, where $A_2=A_1$, 
${\Gamma}_2$ = $\{\tilde{\gamma}_x\}$ and $\tilde{\gamma}_x(x)=\lfloor 2x\rfloor$.
\end{example}

%% file: control-abstraction.tex
\section{Control Abstraction Computation}
\label{self.loop.undec.prop.proof.subsec}
\label{sec:control-abstraction}

As explained in Sect.~\ref{motivations.subsec}, the heaviest computation step
for \qks\ is the computation of the control abstraction. In this section, we
recall the definition of control abstraction, as well as how it is computed by
\qks.

Control abstraction (Def. \ref{def:ctr-abs}) models how a DTLHS 
${\cal H}$ is \emph{seen} from the control software after AD
conversions.  Since QFC control rests on AD conversion we must be
careful not to drive the plant outside the bounds in which AD
conversion works correctly. This leads to the definition of
\emph{admissible action} (Def. \ref{def:safe-action}). 
Intuitively, an action is admissible in a state 
if it never drives the system outside of its 
admissible region.

\begin{definition}[Admissible actions]
\label{def:safe-action}
Let ${\cal H} = (X, U, Y, N)$ be a DTLHS 
and ${\cal Q}=(A, \Gamma)$ be a quantization 
for ${\cal H}$.
An action $u \in A_U$ is $A$-\emph{admissible} in
$s \in A_X$ 
if for all $s'$, 
$(\exists y \in A_Y: N(s, u, y, s'))$
implies
$s' \in A_X$.
An action $\hat{u} \in \Gamma(A_U)$
is ${\cal Q}$-{\em admissible} in $\hat{s} \in \Gamma(A_X)$
if
for all 
$s \in \Gamma^{-1}(\hat{s})$, 
$u \in \Gamma^{-1}(\hat{u})$, $u$ is $A$-admissible for $s$ in ${\cal H}$.
%
\end{definition}

\sloppy

\begin{definition}[Control abstraction]
  \label{def:ctr-abs}\label{control-abstraction.def}
  Let ${\cal H} = (X, U, Y, N)$ be a DTLHS and ${\cal Q}=(A, \Gamma)$ 
  be a quantization for ${\cal H}$. 
  We say that the LTS $\hat{\cal H} =
  (\Gamma(A_X)$, $\Gamma(A_U)$, $\hat{N})$ 
  is a ${\cal Q}$ \emph{control abstraction} of ${\cal H}$ 
  if its transition relation $\hat{N}$ satisfies 
  the following conditions:

  \begin{enumerate}

  \item
    \label{item:witness}
    Each abstract transition stems from a concrete transition.
    Formally:
    for all 
    $\hat{s}, \hat{s}' \in \Gamma(A_X)$, $\hat{u} \in \Gamma(A_U)$,
    if $\hat{N}(\hat{s}, \hat{u}, \hat{s}')$
    then 
    there exist 
    $s \in \Gamma^{-1}(\hat{s})$,
    $u \in \Gamma^{-1}(\hat{u})$,
    $s' \in \Gamma^{-1}(\hat{s}')$, 
    $y \in A_Y$
    such that 
    $N(s, u, y, s')$.


%

  \item
    \label{item:nonloop}
    Each concrete transition 
    is faithfully represented by an abstract transition, 
    whenever it is not a self loop and its corresponding abstract action is ${\cal Q}$-admissible.
    Formally:
    for all $s,s'\in A_X$, $u\in A_U$ 
    such that $\exists y: N(s, u, y, s')$, 
    if $\Gamma(u)$ is ${\cal Q}$-admissible in $\Gamma(s)$ 
    and $\Gamma(s)\not=\Gamma(s')$ 
    then $\hat{N}(\Gamma(s), \Gamma(u), \Gamma(s'))$.
    %

  \item
    \label{item:loop}
    If there is no upper bound to the length of concrete paths inside the 
    counter-image of an abstract state then there is an abstract self loop.
    Formally:
    for all 
    $\hat{s} \in \Gamma(A_X)$, $\hat{u} \in \Gamma(A_U)$,
    if it exists an infinite run $\pi$ in ${\cal H}$ such that $\forall t\in\N$
    $\pi^{(S)}(t)\in\Gamma^{-1}(\hat{s})$ and $\pi^{(A)}(t)\in\Gamma^{-1}(\hat{u})$
    then
    $\hat{N}(\hat{s}, \hat{u}, \hat{s})$. A self loop $(\hat{s}, \hat{u}, \hat{s})$
    of $\hat{N}$ satisfying the above property is said to be a {\em non-eliminable self
    loop}, and {\em eliminable self
    loop} otherwise.
\end{enumerate}
%
\end{definition}

\fussy

\begin{algorithm}
  \caption[Synthesis: Building control abstractions]{Building control abstractions}
  \label{ctr-abs.alg}\label{symb.alg.rat}
  \begin{algorithmic}[1]
    \REQUIRE
    DTLHS ${\cal H} = (X, U, Y, N)$, quantization ${\cal Q}=(A, \Gamma)$.
    \ENSURE {\fun{minCtrAbs}
    $({\cal H}$, ${\cal Q})$} \label{name.alg.step}
    \STATE $\hat{N} \gets \varnothing$ \label{init.alg.step}
    \FORALL {$\hat{x} \in \Gamma(A_{X})$} \label{forall_s.alg.step}
      \STATE $\hat{N} \gets \fun{minCtrAbsAux}({\cal H}, {\cal Q}, \hat{x},
      \hat{N})$ \label{call_aux.step}
    \ENDFOR
    \STATE {\bf return} $(\Gamma(A_X), \Gamma(A_U), \hat{N})$\label{return.alg.step}
  \end{algorithmic}
\end{algorithm}  

Function \fun{minCtrAbs} in
Alg.~\ref{ctr-abs.alg}, given a quantization ${\cal Q} = (A, \Gamma)$ for a
DTLHS ${\cal H}
= (X, U, Y, N)$, computes a ${\cal Q}$-control abstraction $(\Gamma(A_X),$
$\Gamma(A_U),$ $\hat{N})$ of ${\cal H}$ following
Def.~\ref{control-abstraction.def}. 
Namely, for each abstract state $\hat{x}$
(line~\ref{forall_s.alg.step}) an auxiliary function \fun{minCtrAbsAux} is
called. On its side, function \fun{minCtrAbsAux} (which is detailed in
Alg.~\ref{ctr-abs-aux.alg}) decides which transitions, among the ones starting
from $\hat{x}$, fulfills Def.~\ref{control-abstraction.def}. Such
transitions are added to the current partial control abstraction $\hat{N}$.
The new partial control abstraction $\hat{N}$, extending the input control
abstraction with all transitions
starting from $\hat{x}$ and fulfilling Def.~\ref{control-abstraction.def}, is returned at
step~\ref{return.alg.aux.step} of function \fun{minCtrAbsAux}.
Finally, note that the checks in lines~\ref{check_s_u.alg.step},~\ref{self_loop.alg.step}
and~\ref{check_s_s_prime.alg.step}, and the computation in
line~\ref{overimg.alg.step} are performed by properly defining MILP
problems, which are solved
using known algorithms (available in the GLPK package).

\begin{algorithm}
  \caption[Synthesis: Building control abstractions]{Building control
  abstractions: transitions from a given abstract state}
  \label{ctr-abs-aux.alg}\label{symb.alg.aux.rat}
  \begin{algorithmic}[1]
    \REQUIRE
    DTLHS ${\cal H}$, quantization ${\cal Q}$,
    abstract state $\hat{x}$, partial control abstraction $\hat{N}$.
    \ENSURE {\fun{minCtrAbsAux}
    $({\cal H}$, ${\cal Q}$, $\hat{x}$, $\hat{N})$}\label{name.alg.aux.step}
     \FORALL {$\hat{u} \in \Gamma(A_{U})$} \label{forall_u.alg.step} 
      \IF{$\neg$ \fun{${\cal Q}$-admissible}$({\cal H}, {\cal Q}, \hat{x},
      \hat{u})$}
		\label{check_s_u.alg.step} 
      \STATE 
      {{\bf if} \fun{selfLoop}(${\cal H}$,${\cal Q}$, $\hat{x}$,$\hat{u}$) 
{\bf then} $\hat{N} \gets \hat{N} \cup \{(\hat{x}, \hat{u}, 
\hat{x})\}$}  
      \label{self_loop.alg.step}
	  \STATE ${\cal O}$ $\gets$ \fun{overImg}$({\cal H}, {\cal Q}, \hat{x}, \hat{u})$
	  \label{overimg.alg.step}
      \FORALL {$\hat{x}' \in \Gamma({\cal O})$}
        \label{forall_s_prime.alg.step}
	\IF {
	{$\hat{x} \neq \hat{x}'\land$}$\mbox{\fun{existsTrans}}({\cal H}, {\cal Q}, \hat{x},\hat{u},\hat{x}')$}
\label{check_s_s_prime.alg.step}
        \STATE $\hat{N}\!\gets\!\hat{N} \cup \{(\hat{x}, \hat{u}, \hat{x}')\}$\label{update_t.alg.step}
        \ENDIF
      \ENDFOR
        \ENDIF
     \ENDFOR
    \STATE {\bf return} $\hat{N}$\label{return.alg.aux.step}
  \end{algorithmic}
\end{algorithm}

%% file: parallel-algo.tex
\section{Parallel Synthesis of Control Software}\label{parallel-algo.tex}

In this section we present our novel parallel algorithm for the control
abstraction generation of a given DTLHS. 
Such algorithm is a parallel version 
of the standalone 
Alg.~\ref{ctr-abs.alg}. In this way we significantly improve the performance on the
control
abstraction generation (which is the bottleneck of \qks), thus obtaining a
huge speedup for the whole approach to the synthesis of control software
for DTLHSs.

%
In the following, let ${\cal H} = (X,$ $U,$ $Y,$ $N)$, ${\cal Q} = (A,
\Gamma)$ be, respectively, the DTLHS and the quantization in input to our
algorithm for control abstraction generation. Moreover, let $b$ be the overall number of bits
needed in ${\cal Q}$ to quantize plant states (i.e., $b = \sum_{x \in X} b_x$, where $b_x$ is the number 
of bits for $\gamma_x \in \Gamma$). Finally, let $p$ be the number of
processors 
available for parallel computation.

Our parallel algorithm rests on the observation that all calls to function \fun{minCtrAbsAux} (see Alg.~\ref{ctr-abs-aux.alg}) 
are independent of each other, thus they may be performed by independent processes without communication overhead. 
This observation allows us to
use parallel methods targeting {\em embarrassingly parallel} problems in order to obtain
a significant speedup on the control abstraction generation phase. 
To this aim, we use a Map-Reduce based parallelization technique to design a parallel version 
of Alg.~\ref{ctr-abs.alg}. 
Namely, our parallel computation is designed as follows (see Fig.~\ref{parallel-algo.fig} for an example). 

\begin{enumerate}

	\item A {\em master} process assigns ({\em maps}) the computations needed
for an abstract state $\hat{x}$ (i.e., the execution of a call to function
\fun{minCtrAbsAux} of Alg.~\ref{ctr-abs-aux.alg}) to one of $p$ computing processes
({\em workers}, enumerated from 1 to $p$). This is done in a way so that each worker  approximately
handles $\frac{|\Gamma(A_X)|}{p}$ abstract states, thus balancing the parallel
workload.  Namely, abstract states are enumerated from $1$ to $2^b$, and
abstract state $i$ is assigned to worker $1 + ((i - 1)$ mod $p)$. We
denote with  $\Gamma^{(i, p)}(A_X) \subseteq \Gamma(A_X)$ the set of abstract
states mapped to worker $i$ out of $p$ available workers. Note that worker $i$ may locally decide which abstract states are in 
$\Gamma^{(i, p)}(A_X)$ by only knowing $i$ and $p$ (together with the overall input ${\cal H}$ and ${\cal Q}$).
This allows us to avoid sending to each worker the
explicit list of abstract states it has to work on, since it is sufficient that the master
sends $i$ and $p$ (plus ${\cal H}$ and ${\cal Q}$) to worker $i$. 

	\item Each worker {\em works} on its abstract states partition $\Gamma^{(i,
p)}(A_X)$, by calling \fun{minCtrAbsAux} for each abstract state in such partition. 
Once worker $i$ has completed its task (i.e., all abstract states in $\Gamma^{(i, p)}(A_X)$ have been considered),
a local (partial) control abstraction $\hat{N}_i$ is obtained, which is sent
back to the master. 

	\item The master collects the local control  abstractions coming
from the workers and composes ({\em reduces}) them 
in order to obtain the desired complete control abstraction for ${\cal H}$. Note
that, as in embarrassingly parallel tasks, communication only takes place at the
beginning and at the end of local computations. 

\end{enumerate}

\begin{algorithm}
  \caption[Synthesis: Building control abstractions in parallel]{Building
  control abstractions in parallel: master process}
  \label{par-ctr-abs.master.alg}
  \begin{algorithmic}[1]
    \REQUIRE
    DTLHS ${\cal H}$, quantization ${\cal Q}$, workers number $p$ \label{also_i_p.master.step}
    \ENSURE {\fun{minCtrAbsMaster}
    $({\cal H}$, ${\cal Q}$, $p)$}
    \FORALL{$i \in \{1, \ldots, p\}$}\label{map.forall.master.step}
    \STATE create a worker and send ${\cal H}$, ${\cal Q}$, $i$ and $p$ to it \label{map.master.step}
    \ENDFOR
    \STATE wait to get $\hat{N}_1, \ldots, \hat{N}_{p}$ from workers \label{obdds-receiving.master.step}
    \STATE {\bf return} $(\Gamma(A_X), \Gamma(A_U), \cup_{j=1}^{p}\hat{N}_j)$\label{global-ctrl-abs.2.master.step}
  \end{algorithmic}
\end{algorithm}  


\newcommand{\drawpartgrid}[5]{
  \pgfmathtruncatemacro{\tmpcomp}{(#3 - #1)/#5}
  \foreach \manualgrid in {0, ..., \tmpcomp} {
    \draw[help lines] (#1, #2 + \manualgrid*#5) to (#3, #2 + \manualgrid*#5); 
    \draw[help lines] (#1 + \manualgrid*#5, #2) to (#1 + \manualgrid*#5, #4); 
  }
}


\newcommand{\drawallpart}[6]{
  \pgfmathtruncatemacro{\numcells}{(#3 - #1)*(#4 - #2)}
  \foreach \cell in {1,...,\numcells} {
    \pgfmathtruncatemacro{\p}{mod(\cell-1,#5) + 1}
    \pgfmathtruncatemacro{\x}{mod(\cell-1,(#3 - #1))}
    \pgfmathtruncatemacro{\y}{(\cell-1)/((#3 - #1))}
    \draw (#1 + \x+.5*#6, #2 + \y+.5*#6) node {\p};
  }
}

\newcommand{\drawrealpar}[4]{
  \begin{scope}[very thick]
    \ifnum #4=3
      \draw (0*#3 + #1, 0*#3 + #2) rectangle (0*#3 + #1 + #3, 0*#3 + #2 + #3);
      \draw (2*#3 + #1, 0*#3 + #2) rectangle (2*#3 + #1 + #3, 0*#3 + #2 + #3);
      \draw (3*#3 + #1, 0*#3 + #2) rectangle (3*#3 + #1 + #3, 0*#3 + #2 + #3);
      \draw (0*#3 + #1, 1*#3 + #2) rectangle (0*#3 + #1 + #3, 1*#3 + #2 + #3);
      \draw (0*#3 + #1, 2*#3 + #2) rectangle (0*#3 + #1 + #3, 2*#3 + #2 + #3);
      \draw (3*#3 + #1, 2*#3 + #2) rectangle (3*#3 + #1 + #3, 2*#3 + #2 + #3);
      \draw (2*#3 + #1, 3*#3 + #2) rectangle (2*#3 + #1 + #3, 3*#3 + #2 + #3);
    \else
      \ifnum #4=0
	\draw (0*#3 + #1, 0*#3 + #2) rectangle (0*#3 + #1 + #3, 0*#3 + #2 + #3);
	\draw (2*#3 + #1, 0*#3 + #2) rectangle (2*#3 + #1 + #3, 0*#3 + #2 + #3);
	\draw (3*#3 + #1, 0*#3 + #2) rectangle (3*#3 + #1 + #3, 0*#3 + #2 + #3);
      \else
	\ifnum #4=1
	  \draw (0*#3 + #1, 1*#3 + #2) rectangle (0*#3 + #1 + #3, 1*#3 + #2 + #3);
	  \draw (0*#3 + #1, 2*#3 + #2) rectangle (0*#3 + #1 + #3, 2*#3 + #2 + #3);
	\else
	  \draw (3*#3 + #1, 2*#3 + #2) rectangle (3*#3 + #1 + #3, 2*#3 + #2 + #3);
	  \draw (2*#3 + #1, 3*#3 + #2) rectangle (2*#3 + #1 + #3, 3*#3 + #2 + #3);
	\fi
      \fi
    \fi
    \ifnum #4=3
      \draw (2*#3 + #1 + 0.1*#3, 3*#3 + #2) to (2*#3 + #1 + 0.1*#3, 3*#3 + #2 - 0.5*#3);
      \draw (3*#3 + #1 + 0.9*#3, 2*#3 + #2 + #3) to (3*#3 + #1 + 0.9*#3, 2*#3 + #2 + #3 + 0.5*#3);
      \draw (0*#3 + #1 + #3, 1*#3 + #2 + 0.5*#3) to (0*#3 + #1 + #3 + 0.9*#3, 1*#3 + #2 + 0.5*#3);
      \draw (0*#3 + #1 + #3 + 0.9*#3, 1*#3 + #2 + 0.5*#3) to (0*#3 + #1 + #3 + 0.9*#3, 1*#3 + #2 + 1.5*#3);
      \begin{scope}[->]
	\draw (0*#3 + #1 + #3, 0*#3 + #2 + 0.3*#3) to (2*#3 + #1, 0*#3 + #2 + 0.3*#3);
	\draw (0*#3 + #1 + #3, 0*#3 + #2 + 0.7*#3) to [bend left] (3*#3 + #1 + 0.5*#3, 0*#3 + #2 + #3);
	\draw (0*#3 + #1 + #3 + 0.9*#3, 1*#3 + #2 + 1.5*#3) to (0*#3 + #1 + #3, 2*#3 + #2 + 0.5*#3);
	\draw (2*#3 + #1 + 0.1*#3, 3*#3 + #2 - 0.5*#3) to (3*#3 + #1, 2*#3 + #2 + 0.5*#3);
	\draw (3*#3 + #1 + 0.9*#3, 2*#3 + #2 + #3 + 0.5*#3) to (2*#3 + #1 + #3, 3*#3 + #2 + 0.5*#3);
      \end{scope}
    \else
      \ifnum #4=0
	\begin{scope}[->]
	  \draw (0*#3 + #1 + #3, 0*#3 + #2 + 0.3*#3) to (2*#3 + #1, 0*#3 + #2 + 0.3*#3);
	  \draw (0*#3 + #1 + #3, 0*#3 + #2 + 0.7*#3) to [bend left] (3*#3 + #1 + 0.5*#3, 0*#3 + #2 + #3);
	\end{scope}
      \else
	\ifnum #4=1
	  \draw (0*#3 + #1 + #3, 1*#3 + #2 + 0.5*#3) to (0*#3 + #1 + #3 + 0.9*#3, 1*#3 + #2 + 0.5*#3);
	  \draw (0*#3 + #1 + #3 + 0.9*#3, 1*#3 + #2 + 0.5*#3) to (0*#3 + #1 + #3 + 0.9*#3, 1*#3 + #2 + 1.5*#3);
	  \begin{scope}[->]
	    \draw (0*#3 + #1 + #3 + 0.9*#3, 1*#3 + #2 + 1.5*#3) to (0*#3 + #1 + #3, 2*#3 + #2 + 0.5*#3);
	  \end{scope}
	\else
	  \draw (2*#3 + #1 + 0.1*#3, 3*#3 + #2) to (2*#3 + #1 + 0.1*#3, 3*#3 + #2 - 0.5*#3);
	  \draw (3*#3 + #1 + 0.9*#3, 2*#3 + #2 + #3) to (3*#3 + #1 + 0.9*#3, 2*#3 + #2 + #3 + 0.5*#3);
	  \begin{scope}[->]
	    \draw (2*#3 + #1 + 0.1*#3, 3*#3 + #2 - 0.5*#3) to (3*#3 + #1, 2*#3 + #2 + 0.5*#3);
	    \draw (3*#3 + #1 + 0.9*#3, 2*#3 + #2 + #3 + 0.5*#3) to (2*#3 + #1 + #3, 3*#3 + #2 + 0.5*#3);
	  \end{scope}
	\fi
      \fi
    \fi
  \end{scope}
}

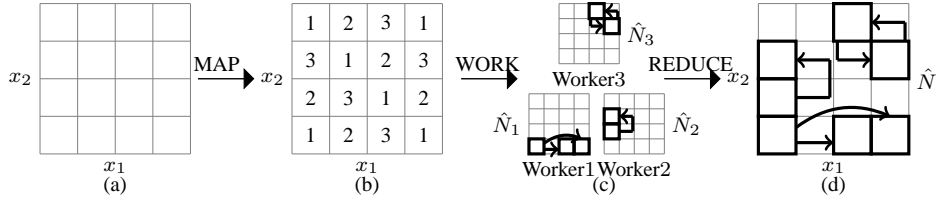
\begin{figure}
  \begin{center}
    \begin{tikzpicture}[scale=.5]
      \scriptsize
      \draw node[below] at (2,0) {$x_1$};
      \draw node[below=0.2] at (2,0) {(a)};
      \draw node[left] at (0,2) {$x_2$};
      \drawpartgrid{0}{0}{4}{4}{1}
      \draw node[above] at (4.7,2) {MAP};
      \draw[-triangle 90,fill=black]  (4.2,2) -- (5.7,2);
      \draw node[below] at (8.7,0) {$x_1$};
      \draw node[left] at (6.7,2) {$x_2$};
      \draw node[below=0.2] at (8.7,0) {(b)};
      \drawpartgrid{6.7}{0}{10.7}{4}{1}
      \drawallpart{6.7}{0}{10.7}{4}{3}{1}
      \draw node[above] at (11.9,2) {WORK};
      \draw[-triangle 90,fill=black]  (11.2,2) -- (12.7,2);
      \draw node[below] at (13.8,0) {Worker$1$};
      \draw node[below] at (15.8,0) {Worker$2$};
      \draw node[below] at (14.6,2.4) {Worker$3$};
      \draw node[below=0.2] at (15, 0) {(c)};
      \draw node[left] at (13,0.8) {$\hat{N}_1$};
      \draw node[right] at (16.6,0.8) {$\hat{N}_2$};
      \draw node[right] at (15.4,3.2) {$\hat{N}_3$};
      \drawpartgrid{13}{0}{14.6}{1.6}{0.4}
      \drawrealpar{13}{0}{0.4}{0}
      \drawpartgrid{15}{0}{16.6}{1.6}{0.4}
      \drawrealpar{15}{0}{0.4}{1}
      \drawpartgrid{13.8}{2.4}{15.4}{4}{0.4}
      \drawrealpar{13.8}{2.4}{0.4}{2}
      \draw node[above] at (17.3,2) {REDUCE};
      \draw[-triangle 90,fill=black]  (16.6,2) -- (18.1,2);
      \draw node[below] at (21.1,0) {$x_1$};
      \draw node[left] at (19.1,2) {$x_2$};
      \draw node[below=0.2] at (21.1,0) {(d)};
      \draw node[right] at (23.1,2) {$\hat{N}$};
      \drawpartgrid{19.1}{0}{23.1}{4}{1}
      \drawrealpar{19.1}{0}{1}{3}
    \end{tikzpicture}
  \end{center}
\caption{Example of execution of the parallel algorithm using 3 workers 
on a DTLHS ${\cal H} = (X, U, Y, N)$ and a quantization ${\cal Q}$ for
${\cal H}$ s.t.  $X = [x_1, x_2]$ and ${\cal Q}$  discretizes both $x_1, x_2$
with two bits. 
In (a) the starting point is shown,
where each cell corresponds to an abstract state.
In (b), function \fun{minCtrAbsMaster} maps the workload among the 3 workers (abstract states labeled with
$i \in [3]$ are handled by worker $i$). In (c) each worker $i$ computes its local control abstraction $\hat{N}_i$, which is 
assumed to have the shown transitions only.
Finally, in (d) the master rejoins the local control
abstractions in order to get the final one, i.e. $\hat{N}$.}\label{parallel-algo.fig}
\end{figure}

Our parallel algorithm is described in
Algs.~\ref{par-ctr-abs.master.alg} (for the master)
and~\ref{par-ctr-abs.worker.alg} (for workers). 

\begin{algorithm}
  \caption[Synthesis: Building control abstractions in parallel]{Building
  control abstractions in parallel: worker processes}
  \label{par-ctr-abs.worker.alg}
  \begin{algorithmic}[1]
    \REQUIRE
    DTLHS ${\cal H} = (X, U, Y, N)$, quantization ${\cal Q}=(A, \Gamma)$,
    index $i$, workers number $p$ \label{also_i_p.worker.step}
    \ENSURE {\fun{parMinCtrAbs}
    $({\cal H}$, ${\cal Q}$, $i$, $p)$}
    \STATE $\hat{N}_i \gets \varnothing$
    \FORALL {$\hat{x} \in \Gamma^{(i, p)}(A_{X})$} \label{par-for-loop.worker.step}
      \STATE $\hat{N}_i \gets \fun{minCtrAbsAux}({\cal H}, {\cal Q}, \hat{x},
      \hat{N}_i)$
    \ENDFOR
    \STATE send $\hat{N_i}$ to the master\label{obdds-sending.worker.step}
  \end{algorithmic}
\end{algorithm}  

\subsection{Implementation with MPI}\label{mpi.subsec}


We actually implemented Algs.~\ref{par-ctr-abs.master.alg}
and~\ref{par-ctr-abs.worker.alg} in \pqks\ by using MPI (Message Passing
Interface, see~\cite{mpi.book}). Since MPI is widely used, this allows us to run \pqks\ on nearly all
computer clusters. Note that in MPI all
computing processes execute the same program, each one knowing its rank
$i$ and the overall number of computing processes $p$ (SPMD paradigm). Thus
lines~\ref{map.forall.master.step}--\ref{map.master.step} of
Alg.~\ref{par-ctr-abs.master.alg} are directly implemented by the MPI framework.
Moreover, in our implementation the master is not a separate node, but it actually
performs like a worker while waiting for local control abstractions from
(other) workers. Local control abstraction from other workers are collected once
the master local control abstraction has been completed. This allows us to use $p$ nodes instead of $p + 1$.



Note that lines~\ref{obdds-receiving.master.step}
and~\ref{obdds-sending.worker.step} of, respectively,
Algs.~\ref{par-ctr-abs.master.alg} and~\ref{par-ctr-abs.worker.alg} require workers to send their local
control abstraction to the master. Being control abstractions
represented as OBDDs ({\em Ordered Binary Decision Diagrams}~\cite{Bry86}), which are sparse data structures, this step may be
difficult to be implemented with a call to MPI\_Send (as it is usually done
in MPI programs), which is designed for contiguous
data. In our experiments, workers use known algorithms (implemented in the CUDD package) to
efficiently dump the OBDD representing their local control abstraction on the
shared filesystem (current MPI implementations are typically based on a shared filesystem). 
Then each computing process calls MPI\_Barrier, 
in order to synchronize all workers with the master.
After this, the master node collects local control abstraction from workers,
by reloading them from the shared filesystem, in order to build the final global one. Consequently, when
presenting experimental results in Sect.~\ref{expres.tex}, we include I/O time
in communication time. Note that communication based on shared
filesystem is very common also in Map-Reduce native
implementations like Hadoop~\cite{mapreduce.book}.



Finally, we note that Algs.~\ref{par-ctr-abs.master.alg} and~\ref{par-ctr-abs.worker.alg} 
may conceptually be implemented on multithreaded
systems with shared memory. 
%
However, in our implementation we use GLPK 
as external library to
solve MILP problems required in
computations inside function \fun{minCtrAbsAux} (see Alg.~\ref{symb.alg.aux.rat}).
Since GLPK is not thread-safe, we may
not implement Algs.~\ref{par-ctr-abs.master.alg} 
and~\ref{par-ctr-abs.worker.alg} on multithreaded shared memory
systems.


%% file: expres.tex
\section{Experimental Results}\label{expres.tex}

%

We implement functions \fun{minCtrAbsMaster} and \fun{parMinCtrAbs} of Algs.~\ref{par-ctr-abs.master.alg} 
and~\ref{par-ctr-abs.worker.alg} in
C programming language using the  CUDD 
package for OBDD based
computations and the GLPK 
package for MILP problems solving, and
MPI for the parallel setting and communication. The resulting tool, \pqks\ 
({\em Parallel QKS}),  extends the tool \qks~\cite{tosem13} by replacing function
\fun{minCtrAbs} of Alg.~\ref{ctr-abs.alg} with function \fun{minCtrAbsMaster} of Alg.~\ref{par-ctr-abs.master.alg}.
%

\sloppy 

In this section we present experimental results obtained by using \pqks\ 
on two meaningful and challenging examples for the automatic synthesis of correct-by-construction 
control software, namely the inverted pendulum and multi-input buck 
DC-DC converter. 
In such experiments, we show the gain of 
the parallel approach with respect to the serial algorithm, also providing
standard measures such as communication and I/O time.

\fussy 

This section is organized as follows. In Sects.~\ref{sect:invpend} and ~\ref{sect:buck}
we will present 
the inverted pendulum and the multi-input 
buck DC-DC converter, on which our experiments focus.
In Sect.~\ref{sect:expressettings} we give the details of the experimental 
setting, and finally, in Sect.~\ref{sect:expresdiscussion}, we discuss 
experimental results.

\input{inverted-pendulum.tex}

\input{multi-input-buck-dc-dc-model.tex}

\input{expres-setting.tex}

\input{expres-discussion.tex}

%% file: inverted-pendulum.tex
\subsection{The Inverted Pendulum Case Study} 
\label{sect:invpend}

\begin{figure}
  \centering
  \includegraphics[scale=0.2]{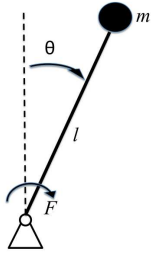}
  \caption{Inverted pendulum with stationary pivot point.}
  \label{fig:invpend}
\end{figure}

The inverted pendulum~\cite{KB94} (see Fig.~\ref{fig:invpend}) is modeled by taking the angle $\theta$ and the angular velocity $\dot{\theta}$ as 
state variables. The input of the system is the torquing force $u\cdot F$, that can influence 
the velocity in both directions. Here, the variable $u$ models the direction and the 
constant $F$ models the intensity of the force. 
Differently from~\cite{KB94}, we consider the problem of finding a discrete controller, whose decisions may be  only ``apply the force clockwise'' ($u=1$), ``apply the force counterclockwise'' ($u=-1$)'', or ``do nothing'' ($u=0$). 
The behavior of the system depends on the pendulum mass $m$, the length of the 
pendulum $l$, and the gravitational acceleration $g$. Given such parameters, 
the motion of the system is described by the differential equation 
$\ddot{\theta} = \dfrac{g}{l} \sin \theta + \dfrac{1}{m l^2} u F$, which may be normalized and discretized in the following
transition relation (being $T$ the sampling time constant, $x_1 = \theta$ and $x_2 = \dot{\theta}$):
$N(x_1, x_2, u, x_1', x_2') \equiv (x'_1 = x_1 + T x_2)\,\land\,(x'_2 = x_2 + T {g \over l} \sin x_1 + T {1 \over m l^2} u F)$.
Such transition relation is not linear, 
as it contains the function $\sin x_1$. 
A linear model can be found by under- and over-approximating the non-linear function 
$\sin x$ on different intervals for $x$. Namely, we may proceed as follows~\cite{AMMSTcdc12}.
First of all, in order to exploit sinus periodicity, we consider the equation $x_1 = 2 \pi y_{k} +
y_{\alpha}$, where $y_{k}$ represents the period in which $x_1$ lies and
$y_{\alpha} \in [-\pi, \pi]$\footnote{In this section we write $\pi$ for a rational approximation of it.} represents the actual $x_1$ inside a given period.
Then, we partition the interval $[-\pi, \pi]$ in four intervals: 
$I_1$ $=$ $\left[-\pi, -\dfrac{\pi}{2}\right]$, 
$I_2$ $=$ $\left[-\dfrac{\pi}{2},0\right]$, $I_3$ $=$ $\left[0, \dfrac{\pi}{2}\right]$, 
$I_4$ $=$ $\left[\dfrac{\pi}{2},\pi\right]$. 
In each interval $I_i$ ($i\in[4]$), we consider two linear functions $f_i^+(x)$ and 
and $f_i^{-}(x)$, such that for all $x\in I_i$, we have that $f_i^-(x)\leq \sin x\leq f_i^+(x)$.
As an example, $f_1^{+}(y_{\alpha})$ $=$ $-0.637 y_{\alpha} - 2$ and 
$f_1^{-}(y_{\alpha})$ $=$ $-0.707 y_{\alpha} - 2.373$. 

Let us consider the set of fresh continuous variables $Y^r$ $=$ $\{y_\alpha,$ $y_{\sin}\}$ 
and the set of fresh discrete variables $Y^d=\{y_k, y_q,$ $y_1, y_2,$ $y_3, y_4\}$, being $y_1, \ldots, y_4$ boolean variables. 
The DTLHS model ${\cal I}_F$ for the inverted pendulum is the tuple $(X,U,$ $Y,N)$, 
where $X=\{x_1, x_2\}$ is the set of continuous state variables, 
$U=\{u\}$ is the set of input variables, $Y=Y^r\,\cup\, Y^d$ is the set of auxiliary variables, and the transition 
relation $N(X,U,Y,X')$ is the following guarded predicate: 
\vspace{-.25cm}
\[
\begin{array}{l}
(x'_1 = x_1 + 2\pi y_q + T x_2)	\,\land\, (x'_2 = x_2 + T\dfrac{g}{l} y_{\sin} + T\dfrac{1}{m l^2} u F)
\\[\medskipamount]
\hspace*{.3cm} \land \,\bigwedge_{i\in[4]} y_i\rightarrow f_i^-(y_\alpha)\leq y_{\sin}\leq f_i^+(y_\alpha)
\\[\medskipamount]
\hspace*{.3cm}\land \, \bigwedge_{i\in[4]} y_i\rightarrow y_\alpha \in I_i 
\land \sum_{i\in[4]} y_i \geq 1
\\[\medskipamount]
\hspace*{.3cm}\land \; x_1=2\pi y_k + y_\alpha \;\land\; -\pi \leq x_1' \leq \pi
\end{array}
\]
Overapproximations of the system behaviour increase system nondeterminism.
Since ${\cal I}_F$ dynamics overapproximates the dynamics of the non-linear model,  
the controllers that we synthesize are inherently {\em robust}, that is they meet 
the given closed loop requirements 
{\em notwithstanding} nondeterministic small \emph{disturbances} such as
variations in the plant parameters.
Tighter overapproximations of non-linear functions makes finding a controller easier, 
whereas coarser overapproximations makes controllers more robust.

The typical goal for the inverted pendulum
is to turn the pendulum steady 
to the upright position, starting from any possible initial position, within a given speed interval.

%% file: multi-input-buck-dc-dc-model.tex
\subsection{The Multi-input Buck DC-DC Converter Case Study}
\label{sect:buck}


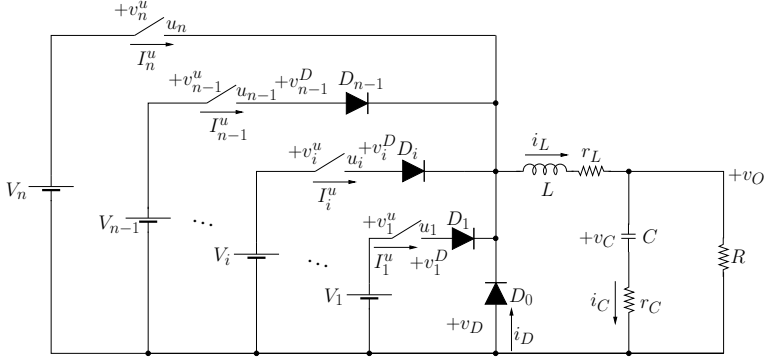
\begin{figure}
  \scalebox{0.3}{\input{multi-input-buck.pstex_t}}
  \caption{Multi-input Buck DC-DC converter.}
  \label{fig:mibdcdc}
\end{figure}

The {\em multi-input} buck DC-DC converter \cite{multin-buck-dcdc-2010} in Fig.~\ref{fig:mibdcdc} is a mixed-mode analog
circuit converting the DC input voltage ($V_i$ in Fig.~
\ref{fig:mibdcdc}) to a desired DC output voltage ($v_O$ in Fig.~
\ref{fig:mibdcdc}).
As an example, buck DC-DC converters are used off-chip to scale down
the typical laptop battery voltage (12-24) to the just few volts
needed by the laptop processor (e.g. \cite{fuzzy-dc-dc-1996}) as well
as on-chip to support \emph{Dynamic Voltage and Frequency Scaling}
(DVFS) in multicore processors
(e.g. \cite{gigascale-integration-07}).
Because of its widespread use,
control schemas for buck DC-DC converters have been widely studied
(e.g. see
\cite{gigascale-integration-07,fuzzy-dc-dc-1996}). 
The typical software based approach (e.g. see \cite{fuzzy-dc-dc-1996}) is to control
the switches $u_1, \ldots, u_n$ in Fig.~\ref{fig:mibdcdc} (typically implemented with a
MOSFET) with a microcontroller.

In such a converter (Fig.~\ref{fig:mibdcdc}),
there are $n$ power supplies with voltage values $V_1, \ldots, V_n$, $n$
switches with voltage values $v_1^u, \ldots, v_n^u$ and current values $I_1^{u},
\ldots, I_n^{u}$, and $n$ input diodes $D_0, \ldots, D_{n - 1}$ with voltage
values $v_0^D, \ldots, v_{n - 1}^D$ and current $i_0^D, \ldots, i_{n - 1}^D$ (in
the following, we will write $v_D$ for $v_0^D$ and $i_D$ for $i_0^D$). 

The circuit state variables are $i_L$ and $v_C$.  However we can
also use the pair $i_L$, $v_O$ as state variables in the DTLHS 
model since there is a linear relationship between $i_L$, $v_C$ and
$v_O$, namely: $v_O \; = \; \frac{r_C R}{r_C + R} i_L + \frac{R}{r_C
  + R} v_C$.
We model the $n$-input buck DC-DC converter with the DTLHS ${\cal B}_n$ = ($X$,
$U$, $Y$, $N$), with $X = [i_L$, $v_O]$, $U = [u_1$, $\ldots$, $u_n]$, $Y =
[v_D$, $v_1^D$, $\ldots, v_{n - 1}^{D}$, $i_D$, $I_1^{u}$, $\ldots$, $I_n^{u}$,
$ v_1^u$, $\ldots$, $v_n^u]$. 

Finally, the transition relation $N$, depending on variables in $X$, $U$ and $Y$
(as well as on circuit parameters $V_i$, $R$, $r_L$, $r_C$, $L$ and $C$), 
may be derived from simple circuit analysis~\cite{AMMSTemsoft12}.
Namely, we have the following equations:
\[
\vspace{-.20cm}
\begin{array}{rcl}
  \dot{i}_L & = & a_{1,1}i_L + a_{1,2} v_{O} + a_{1,3}v_D \\[\medskipamount]
  \dot{v}_{O} & = & a_{2,1}i_L + a_{2,2}v_{O} + a_{2,3}v_D
\\[\smallskipamount] 
\end{array}
\]
where the coefficients $a_{i, j}$ depend on the circuit parameters $R$, $r_L$, $r_C$, $L$ and $C$ in the following way:
$a_{1,1} = -\frac{r_L}{L}$,
$a_{1,2} = -\frac{1}{L}$,
$a_{1,3} = -\frac{1}{L}$,
$a_{2,1} = \frac{R}{r_c + R}[-\frac{r_c r_L}{L} + \frac{1}{C}]$,
$a_{2,2} = \frac{-1}{r_c + R}[\frac{r_c R}{L} + \frac{1}{C}]$,
$a_{2,3} = -\frac{1}{L}\frac{r_c R}{r_c + R}$.
Using a discrete time model with sampling time $T$ (writing $x'$ for
$x(t+1)$) we have:
\begin{eqnarray*}
  {i'_L} & = & (1 + Ta_{1,1})i_L + Ta_{1,2}v_O + Ta_{1,3}v_D  \label{buck:next-il} \\
  {v'_O} & = & Ta_{2,1}i_L + (1 + Ta_{2,2})v_O + Ta_{2,3}v_D. \label{buck:next-vc}
\end{eqnarray*}
The algebraic constraints stemming from the constitutive equations of
the switching elements
are the following:

\begin{center}
\begin{tabular}{lr}

\begin{minipage}{0.45\textwidth}
\small
\begin{eqnarray}
\nonumber
q_0  & \rightarrow & v_D = R_{\rm on} i_D \label{mbuck-begin.eq} \\\nonumber
q_0  & \rightarrow & i_D \geq 0 \\\nonumber
\bigwedge_{i = 1}^{n - 1} q_i & \rightarrow & v_i^D = R_{\rm on} I_i^u \\\nonumber
\bigwedge_{i = 1}^{n - 1} q_i & \rightarrow & I_i^u \geq 0 \\\nonumber
\bigwedge_{j = 1}^{n} u_j & \rightarrow & v_j^u = R_{\rm on} I_j^u\\\nonumber
i_L &  = & i_D + \sum_{i=1}^{n}I_i^{u}
\end{eqnarray}
\end{minipage} &

\begin{minipage}{0.45\textwidth}
\small
\begin{eqnarray}
\nonumber
\bar{q}_0 & \rightarrow &  v_D = R_{\rm off}i_D \\\nonumber
\bar{q}_0 & \rightarrow & v_D \leq 0 \\\nonumber
\bigwedge_{i = 1}^{n - 1} \bar{q}_i & \rightarrow &  v_i^D = R_{\rm off}I_i^u \\\nonumber
\bigwedge_{i = 1}^{n - 1} \bar{q}_i & \rightarrow & v_i^D \leq 0 \\\nonumber
\bigwedge_{j = 1}^{n} \bar{u}_j & \rightarrow & v_j^u = R_{\rm off} I_j^u\\\nonumber
v_D  & = & v_i^u + v_{i}^{D} - V_i \label{Vi.i.eq} \\\nonumber
v_D &  = & v_n^u - V_n\label{mbuck-end.eq}\label{Vn.eq}
\end{eqnarray}
\end{minipage} \\

\end{tabular}
\end{center}

The typical goal for a multi-input buck is to drive $i_L$ and $v_O$ within given goal intervals.

%% file: multi-input-buck.pstex_t
\begin{picture}(0,0)%
\includegraphics{multi-input-buck.pstex}%
\end{picture}%
\setlength{\unitlength}{4144sp}%
\begingroup\makeatletter\ifx\SetFigFont\undefined%
\gdef\SetFigFont#1#2#3#4#5{%
  \reset@font\fontsize{#1}{#2pt}%
  \fontfamily{#3}\fontseries{#4}\fontshape{#5}%
  \selectfont}%
\fi\endgroup%
\begin{picture}(14475,7154)(886,-7314)
\put(15346,-5461){\makebox(0,0)[lb]{\smash{{\SetFigFont{29}{34.8}{\rmdefault}{\mddefault}{\updefault}{\color[rgb]{0,0,0}$R$}%
}}}}
\put(15301,-3796){\makebox(0,0)[lb]{\smash{{\SetFigFont{29}{34.8}{\rmdefault}{\mddefault}{\updefault}{\color[rgb]{0,0,0}$+v_O$}%
}}}}
\put(11566,-4111){\makebox(0,0)[lb]{\smash{{\SetFigFont{29}{34.8}{\rmdefault}{\mddefault}{\updefault}{\color[rgb]{0,0,0}$L$}%
}}}}
\put(11026,-6991){\makebox(0,0)[lb]{\smash{{\SetFigFont{29}{34.8}{\rmdefault}{\mddefault}{\updefault}{\color[rgb]{0,0,0}$i_D$}%
}}}}
\put(901,-4111){\makebox(0,0)[lb]{\smash{{\SetFigFont{29}{34.8}{\familydefault}{\mddefault}{\updefault}{\color[rgb]{0,0,0}$V_n$}%
}}}}
\put(2746,-4786){\makebox(0,0)[lb]{\smash{{\SetFigFont{29}{34.8}{\familydefault}{\mddefault}{\updefault}{\color[rgb]{0,0,0}$V_{n-1}$}%
}}}}
\put(5041,-5416){\makebox(0,0)[lb]{\smash{{\SetFigFont{29}{34.8}{\familydefault}{\mddefault}{\updefault}{\color[rgb]{0,0,0}$V_i$}%
}}}}
\put(7246,-6226){\makebox(0,0)[lb]{\smash{{\SetFigFont{29}{34.8}{\familydefault}{\mddefault}{\updefault}{\color[rgb]{0,0,0}$V_1$}%
}}}}
\put(3511,-1501){\makebox(0,0)[lb]{\smash{{\SetFigFont{29}{34.8}{\rmdefault}{\mddefault}{\updefault}{\color[rgb]{0,0,0}$I^u_n$}%
}}}}
\put(4951,-2851){\makebox(0,0)[lb]{\smash{{\SetFigFont{29}{34.8}{\rmdefault}{\mddefault}{\updefault}{\color[rgb]{0,0,0}$I^u_{n-1}$}%
}}}}
\put(7111,-4246){\makebox(0,0)[lb]{\smash{{\SetFigFont{29}{34.8}{\rmdefault}{\mddefault}{\updefault}{\color[rgb]{0,0,0}$I^u_i$}%
}}}}
\put(3061,-511){\makebox(0,0)[lb]{\smash{{\SetFigFont{29}{34.8}{\rmdefault}{\mddefault}{\updefault}{\color[rgb]{0,0,0}$+v^u_n$}%
}}}}
\put(4096,-781){\makebox(0,0)[lb]{\smash{{\SetFigFont{29}{34.8}{\rmdefault}{\mddefault}{\updefault}{\color[rgb]{0,0,0}$u_n$}%
}}}}
\put(10936,-6226){\makebox(0,0)[lb]{\smash{{\SetFigFont{29}{34.8}{\rmdefault}{\mddefault}{\updefault}{\color[rgb]{0,0,0}$D_0$}%
}}}}
\put(9676,-4696){\makebox(0,0)[lb]{\smash{{\SetFigFont{29}{34.8}{\rmdefault}{\mddefault}{\updefault}{\color[rgb]{0,0,0}$D_1$}%
}}}}
\put(8686,-3301){\makebox(0,0)[lb]{\smash{{\SetFigFont{29}{34.8}{\rmdefault}{\mddefault}{\updefault}{\color[rgb]{0,0,0}$D_i$}%
}}}}
\put(7516,-1906){\makebox(0,0)[lb]{\smash{{\SetFigFont{29}{34.8}{\rmdefault}{\mddefault}{\updefault}{\color[rgb]{0,0,0}$D_{n-1}$}%
}}}}
\put(11386,-3121){\makebox(0,0)[lb]{\smash{{\SetFigFont{29}{34.8}{\rmdefault}{\mddefault}{\updefault}{\color[rgb]{0,0,0}$i_L$}%
}}}}
\put(12376,-3346){\makebox(0,0)[lb]{\smash{{\SetFigFont{29}{34.8}{\rmdefault}{\mddefault}{\updefault}{\color[rgb]{0,0,0}$r_L$}%
}}}}
\put(12331,-5056){\makebox(0,0)[lb]{\smash{{\SetFigFont{29}{34.8}{\rmdefault}{\mddefault}{\updefault}{\color[rgb]{0,0,0}$+v_C$}%
}}}}
\put(13591,-5056){\makebox(0,0)[lb]{\smash{{\SetFigFont{29}{34.8}{\rmdefault}{\mddefault}{\updefault}{\color[rgb]{0,0,0}$C$}%
}}}}
\put(13546,-6361){\makebox(0,0)[lb]{\smash{{\SetFigFont{29}{34.8}{\rmdefault}{\mddefault}{\updefault}{\color[rgb]{0,0,0}$r_C$}%
}}}}
\put(12556,-6316){\makebox(0,0)[lb]{\smash{{\SetFigFont{29}{34.8}{\rmdefault}{\mddefault}{\updefault}{\color[rgb]{0,0,0}$i_C$}%
}}}}
\put(6526,-3346){\makebox(0,0)[lb]{\smash{{\SetFigFont{29}{34.8}{\rmdefault}{\mddefault}{\updefault}{\color[rgb]{0,0,0}$+v^u_i$}%
}}}}
\put(5491,-2131){\makebox(0,0)[lb]{\smash{{\SetFigFont{29}{34.8}{\rmdefault}{\mddefault}{\updefault}{\color[rgb]{0,0,0}$u_{n-1}$}%
}}}}
\put(7696,-3481){\makebox(0,0)[lb]{\smash{{\SetFigFont{29}{34.8}{\rmdefault}{\mddefault}{\updefault}{\color[rgb]{0,0,0}$u_i$}%
}}}}
\put(9676,-6766){\makebox(0,0)[lb]{\smash{{\SetFigFont{29}{34.8}{\rmdefault}{\mddefault}{\updefault}{\color[rgb]{0,0,0}$+v_D$}%
}}}}
\put(6886,-5551){\makebox(0,0)[lb]{\smash{{\SetFigFont{29}{34.8}{\familydefault}{\mddefault}{\updefault}{\color[rgb]{0,0,0}$\ddots$}%
}}}}
\put(4591,-4741){\makebox(0,0)[lb]{\smash{{\SetFigFont{29}{34.8}{\familydefault}{\mddefault}{\updefault}{\color[rgb]{0,0,0}$\ddots$}%
}}}}
\put(8236,-5551){\makebox(0,0)[lb]{\smash{{\SetFigFont{29}{34.8}{\rmdefault}{\mddefault}{\updefault}{\color[rgb]{0,0,0}$I^u_1$}%
}}}}
\put(8956,-5551){\makebox(0,0)[lb]{\smash{{\SetFigFont{29}{34.8}{\rmdefault}{\mddefault}{\updefault}{\color[rgb]{0,0,0}$+v^D_1$}%
}}}}
\put(7921,-3256){\makebox(0,0)[lb]{\smash{{\SetFigFont{29}{34.8}{\rmdefault}{\mddefault}{\updefault}{\color[rgb]{0,0,0}$+v^D_i$}%
}}}}
\put(4141,-1906){\makebox(0,0)[lb]{\smash{{\SetFigFont{29}{34.8}{\rmdefault}{\mddefault}{\updefault}{\color[rgb]{0,0,0}$+v^u_{n-1}$}%
}}}}
\put(6256,-1951){\makebox(0,0)[lb]{\smash{{\SetFigFont{29}{34.8}{\rmdefault}{\mddefault}{\updefault}{\color[rgb]{0,0,0}$+v^D_{n-1}$}%
}}}}
\put(8011,-4741){\makebox(0,0)[lb]{\smash{{\SetFigFont{29}{34.8}{\rmdefault}{\mddefault}{\updefault}{\color[rgb]{0,0,0}$+v^u_1$}%
}}}}
\put(9136,-4831){\makebox(0,0)[lb]{\smash{{\SetFigFont{29}{34.8}{\rmdefault}{\mddefault}{\updefault}{\color[rgb]{0,0,0}$u_1$}%
}}}}
\end{picture}%

%% file: expres-setting.tex
\subsection{Experimental Setting}
\label{sect:expressettings}

All experiments have been carried out on a cluster with 4 nodes and Open MPI
implementation of MPI. Each node contains 4 quad-core 2.83 GHz Intel Xeon E5440
processors. This allows us to run fully parallel experiments by configuring the MPI computation 
to use up to 16 processes per node. In order not to overload each node,
we run maximum 15 processes per
node, thus our upper bound for the number of processes is 60. Finally, as in most clusters, nodes share a common file system.


In the inverted pendulum ${\cal I}_F$ with force intensity $F$, as in~\cite{KB94}, we
set pendulum parameters $l$ and $m$ in such a way that  $\frac{g}{l}=1$ (i.e. $l=g$)
$\frac{1}{ml^2}=1$ (i.e. $m=\frac{1}{l^2}$). As for the admissible region, we set $A_{x_1} =
[-1.1\pi, 1.1\pi]$ (we write $\pi$ for a rational approximation of it) and $A_{x_2} = [-4, 4]$. 

In the multi-input buck DC-DC converter with $n$ inputs ${\cal B}_n$, we set constant parameters as follows:
$L = 2 \cdot 10^{-4}$ H, $r_L = 0.1$ ${\rm \Omega}$, $r_C = 0.1$ ${\rm
\Omega}$,  $R = 5$ ${\rm \Omega}$,  $C = 5 \cdot 10^{-5}$ F,   
and $V_i = 10i$ V for $i\in
[n]$. As for the admissible region, we set $A_{i_L} = [-4, 4]$ and $A_{v_O}
= [-1, 7]$. 

%
As for quantization, we will use an even number of bits $b$, so that each state variable of each case study
is quantized with $\frac{b}{2}$ 
bits.
We recall that the number of abstract states is exactly $2^{b}$.

We run \qks\ and \pqks\ on the inverted pendulum model ${\cal I}_F$ 
with $F = 0.5 N$ (force intensity), 
and on the multi-input buck DC-DC model ${\cal B}_n$, 
with $n = 5$ (number of inputs). 
For the inverted pendulum, we use sampling time $T=0.01$ seconds.
For the multi-input buck, we set $T=10^{-6}$ seconds.
For both systems, we run experiments varying the number of bits $b = 18, 20$ (also $22$ for the inverted pendulum) 
and the number of processors (workers) $p = 1, 10, 20, 30, 40, 50,
60$.

\begin{figure}
\begin{tabular}{cc}
\begin{minipage}{0.45\textwidth}
\hspace*{-3mm}
	\includegraphics[scale=0.45]{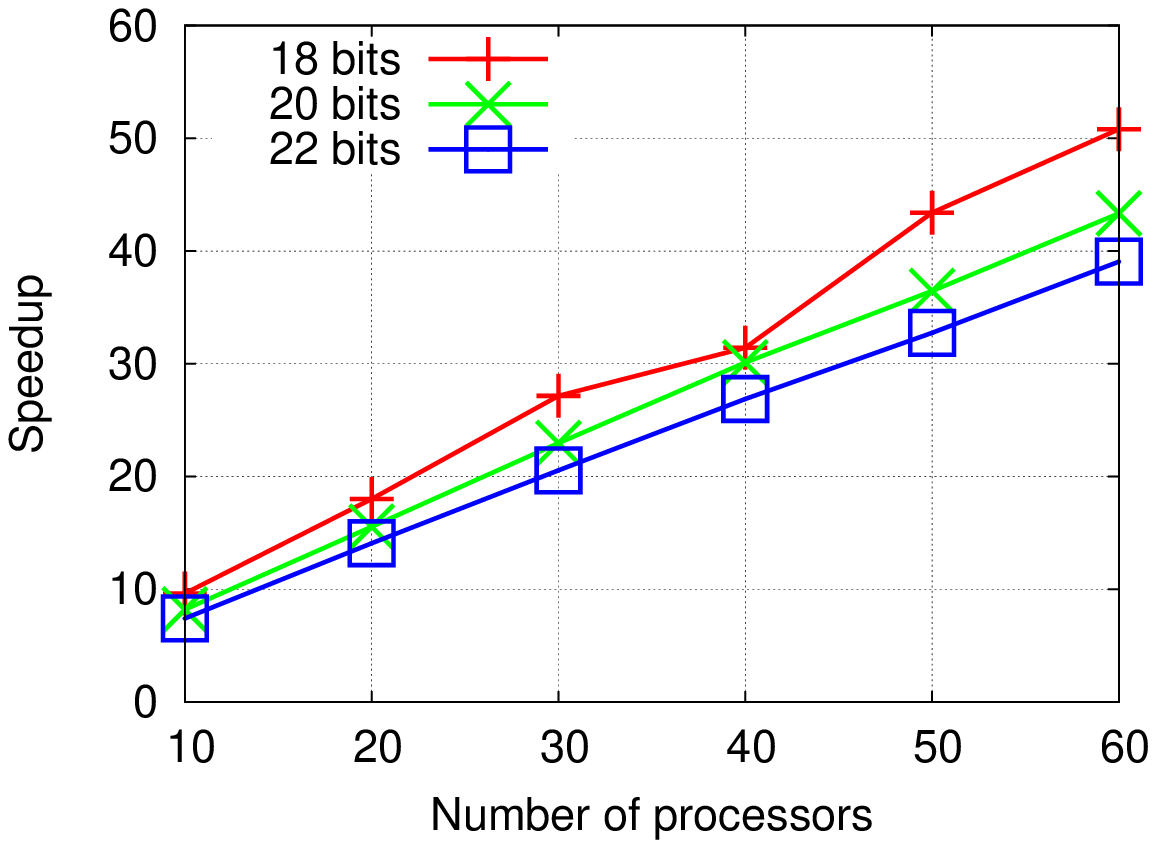}
	\caption{Inverted pendulum: speedup.}
	\label{fig:pendulum_ctrabs_speedup.eps}
\end{minipage}
&
\begin{minipage}{0.45\textwidth}
\hspace*{-3mm}
	\includegraphics[scale=0.45]{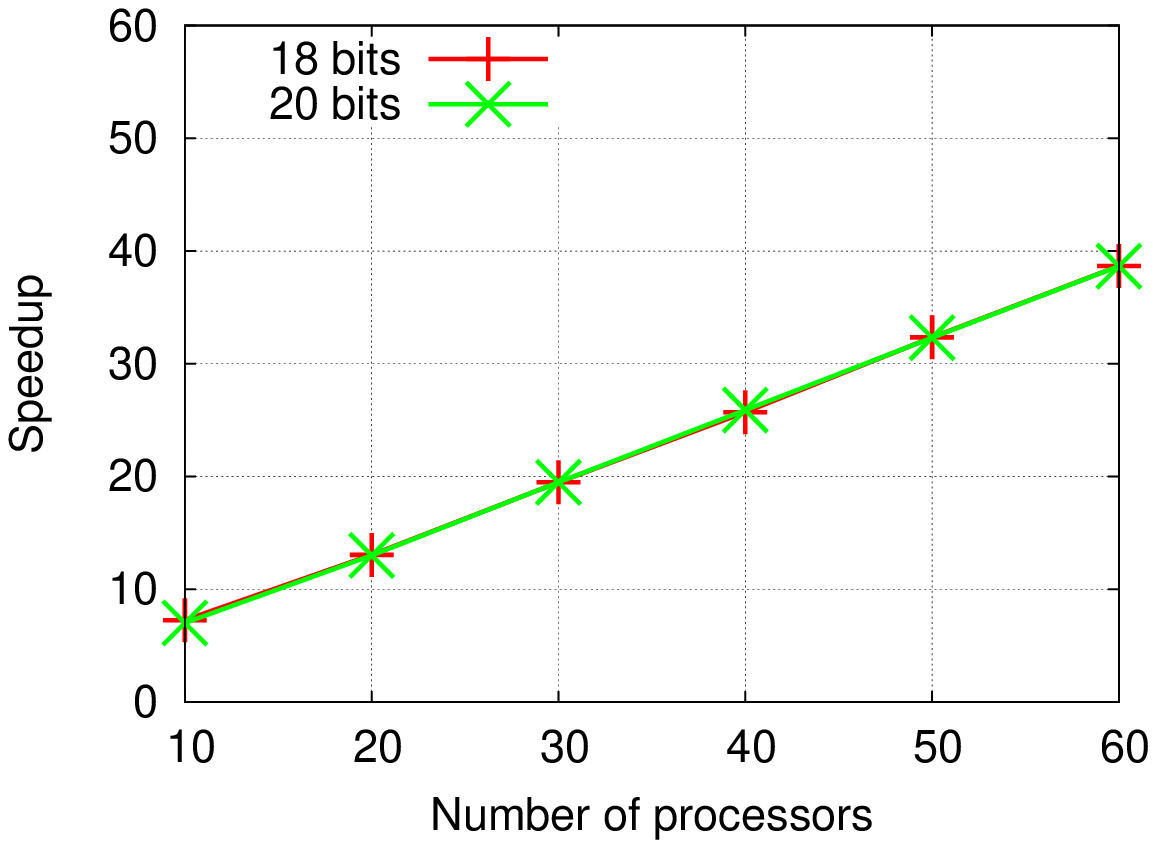}
	\caption{Multi-input buck: speedup.}
	\label{fig:buck_ctrabs_speedup.eps}
\end{minipage}
\end{tabular}
\end{figure}

%% file: expres-discussion.tex
In order to evaluate effectiveness of our approach, we use the following measures: speedup, efficiency, communication time (in seconds) and I/O time (in seconds).
The \emph{speedup} of our approach is represented by the serial CPU time divided by the parallel CPU time, i.e. $\mathrm{Speedup} = {\mathrm{serial} \; \mathrm{CPU} \over \mathrm{parallel} \; \mathrm{CPU}}$.
To evaluate scalability of our approach we define the \emph{scaling efficiency} (or simply \emph{efficiency}) as the percentage ratio between speedup and number of processors $p$, i.e. $\mathrm{Efficiency} = {\mathrm{speedup} \over p} \%$.
In Algs.~\ref{par-ctr-abs.master.alg} 
and~\ref{par-ctr-abs.worker.alg}, the \emph{communication time} consists in the time needed by all workers to send their local control abstraction to the master. In agreement with Sect.~\ref{mpi.subsec}, the communication time is 
increased by the I/O time, that is the overall time spent by processors in input/output activities.

\begin{figure}
\begin{tabular}{cc}
\begin{minipage}{0.45\textwidth}
\hspace*{-3mm}
	\includegraphics[scale=0.45]{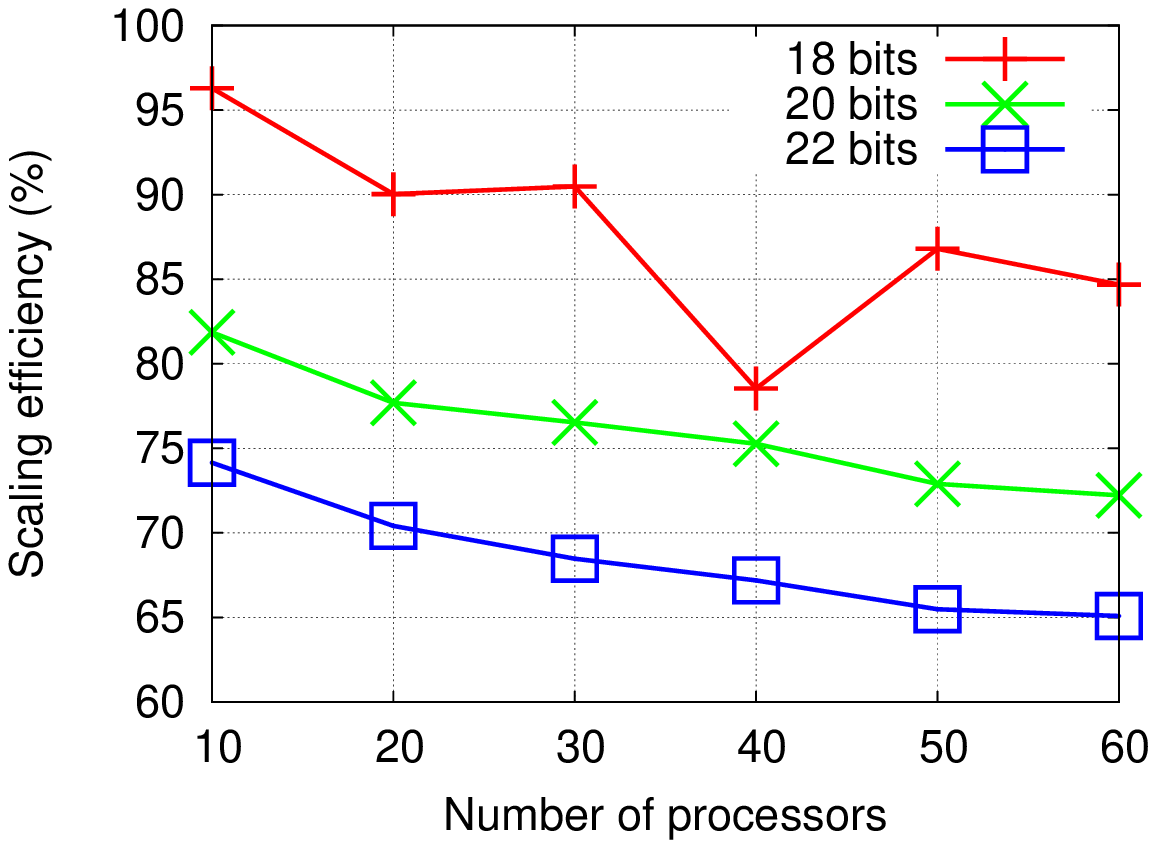}
	\caption{Inverted pendulum: scaling efficiency.}
	\label{fig:pendulum_ctrabs_efficiency.eps}
\end{minipage}
&
\begin{minipage}{0.45\textwidth}
\hspace*{-3mm}
	\includegraphics[scale=0.45]{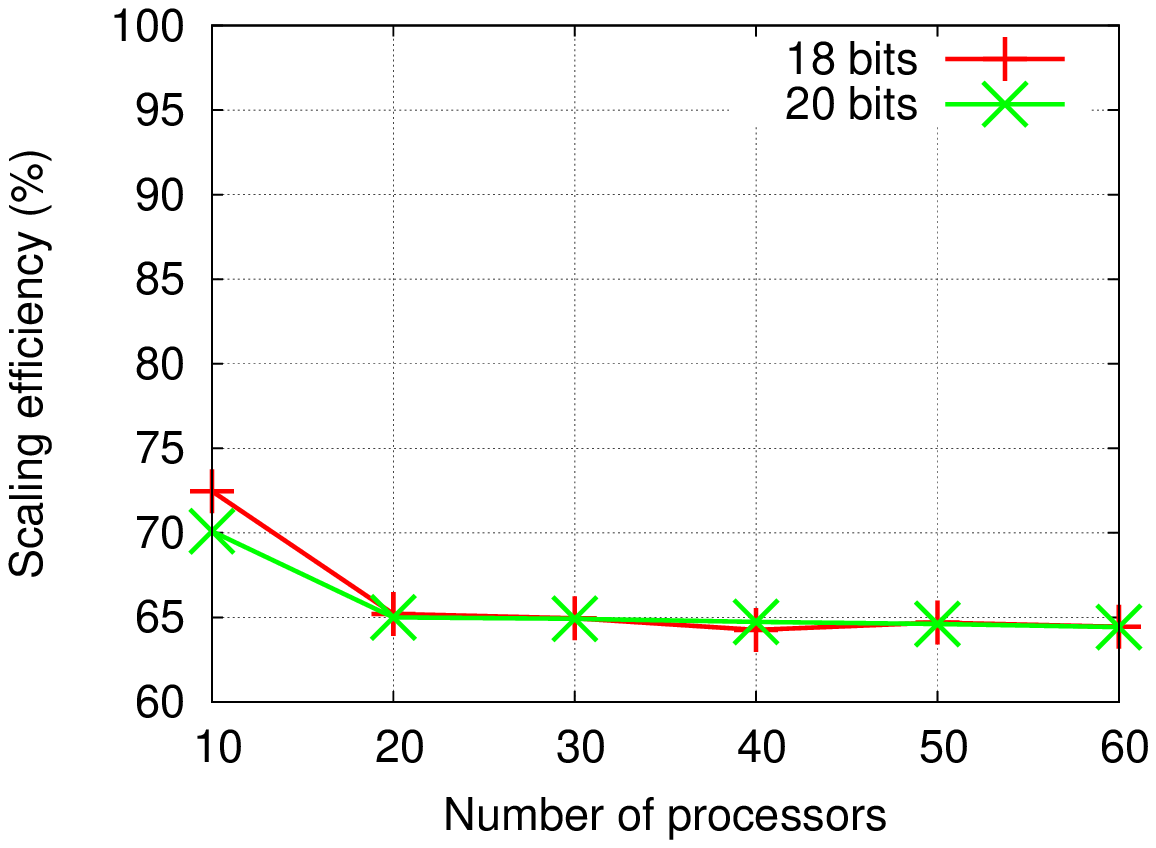}
	\caption{Multi-input buck: scaling efficiency.}
	\label{fig:buck_ctrabs_efficiency.eps}
\end{minipage}
\end{tabular}
\end{figure}

\begin{figure}
\begin{tabular}{cc}
\begin{minipage}{0.45\textwidth}
\hspace*{-3mm}
	\includegraphics[scale=0.45]{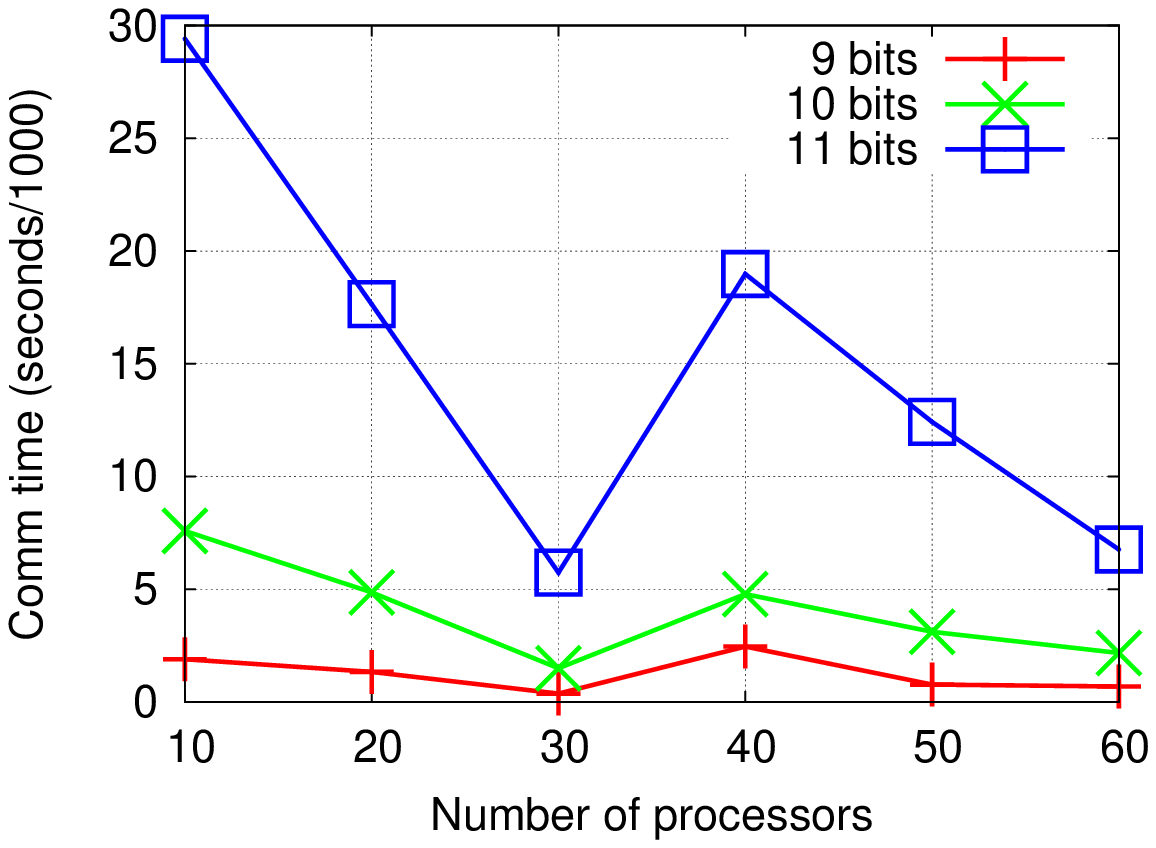}
	\caption{Inverted pendulum: communication time (divided by 1000).}
	\label{fig:pendulum_comm.eps}
\end{minipage}
&
\begin{minipage}{0.45\textwidth}
\hspace*{-3mm}
	\includegraphics[scale=0.45]{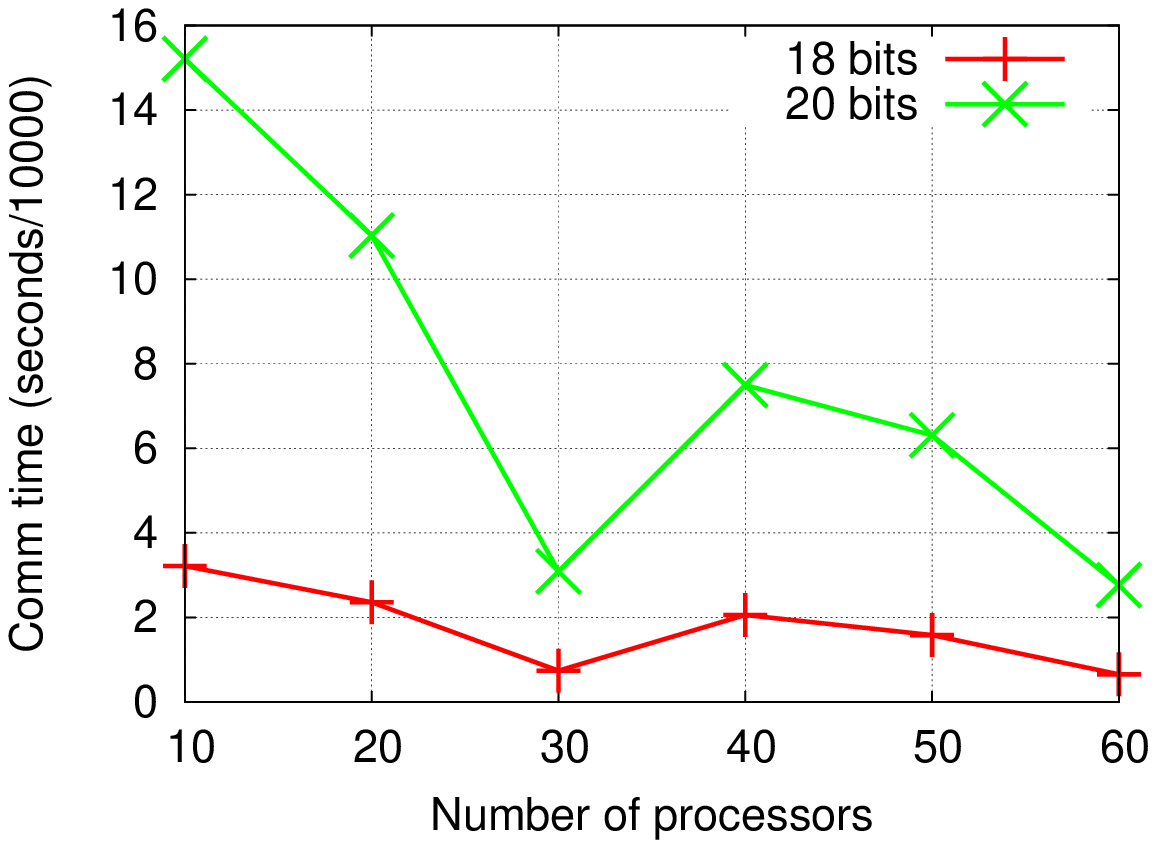}
	\caption{Multi-input buck: communication time (divided by 10000).}
	\label{fig:buck_comm.eps}
\end{minipage}
\end{tabular}
\end{figure}

\begin{figure}
\begin{tabular}{cc}
\begin{minipage}{0.45\textwidth}
\hspace*{-3mm}
	\includegraphics[scale=0.45]{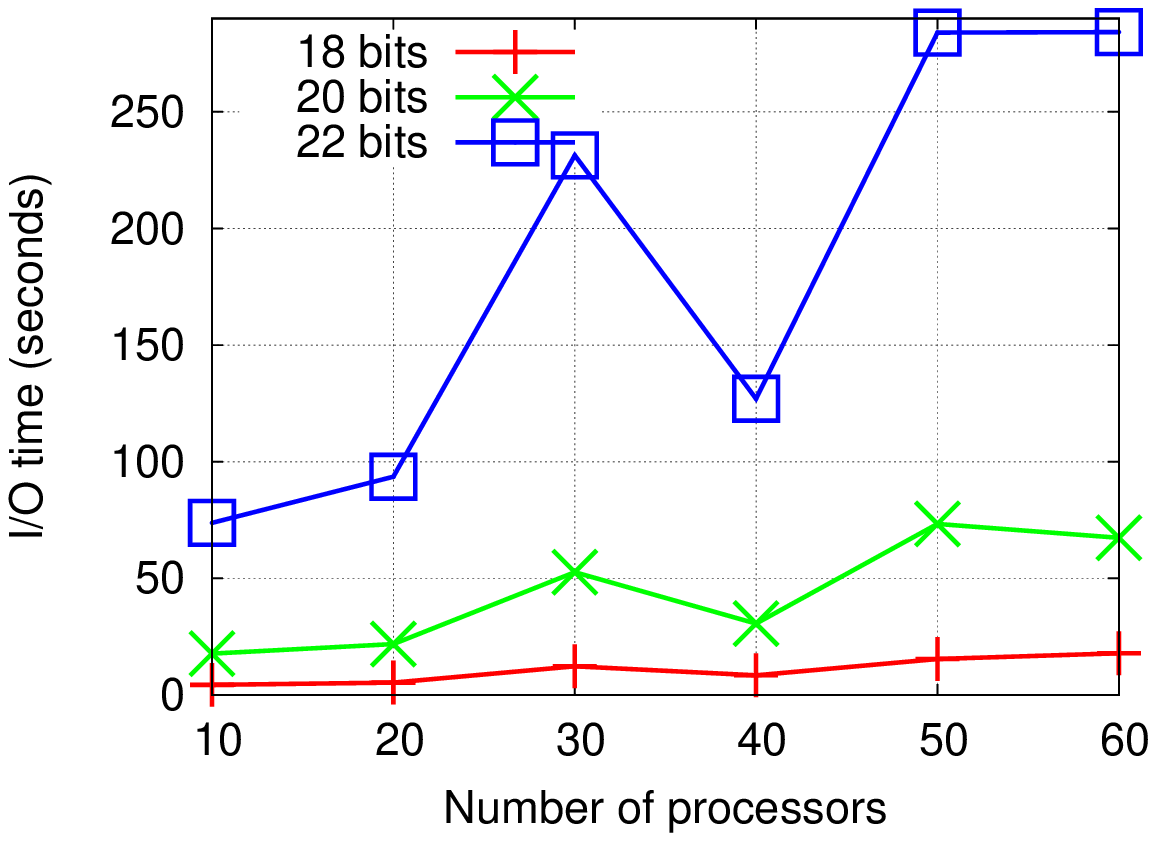}
	\caption{Inverted pendulum: I/O time.}
	\label{fig:pendulum_io.eps}
\end{minipage}
&
\begin{minipage}{0.45\textwidth}
\hspace*{-3mm}
	\includegraphics[scale=0.45]{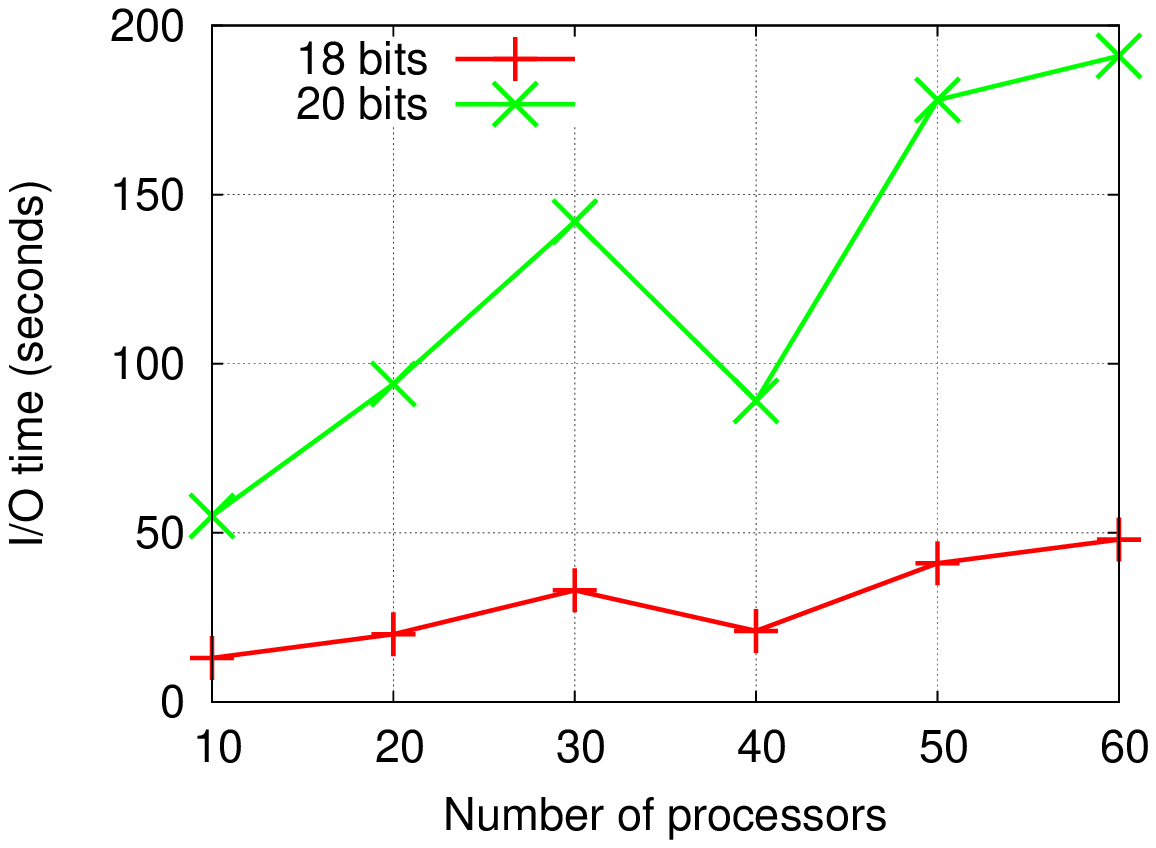}
	\caption{Multi-input buck: I/O time.}
	\label{fig:buck_io.eps}
\end{minipage}
\end{tabular}
\end{figure}

Figs.~\ref{fig:pendulum_ctrabs_speedup.eps},~\ref{fig:pendulum_ctrabs_efficiency.eps},~\ref{fig:pendulum_comm.eps} and~\ref{fig:pendulum_io.eps} show, respectively, the speedup, the scaling efficiency, the communication time (divided by 1000) and the I/O time of Algs.~\ref{par-ctr-abs.master.alg} 
and~\ref{par-ctr-abs.worker.alg} as a function of $p$, for the inverted pendulum with $b = 18, 20, 22$. Analogously, Figs.~\ref{fig:buck_ctrabs_speedup.eps},~\ref{fig:buck_ctrabs_efficiency.eps},~\ref{fig:buck_comm.eps} and~\ref{fig:buck_io.eps} show the same measures (except for the fact that communication time is divided by 10000) for the multi-input buck with $b = 18, 20$.

We also show the absolute values for the experiments with 50 and 60 processors in Tabs.~\ref{tab:invpend} and~\ref{tab:buck}. Tabs.~\ref{tab:invpend} and~\ref{tab:buck} have common columns.
The meaning of such common columns is as follows.
Column \textbf{b} is the number of bits used for quantization.
Column \textbf{QKS} (\textbf{CPU Ctrabs}) reports the execution time in seconds needed by \qks\ to compute the control abstraction (i.e. Alg.~\ref{ctr-abs.alg}).
Columns \textbf{PQKS} report experimental values for \pqks.
Namely, column $p$ shows the number of processors, column \textbf{CPU Ctrabs} reports the execution time in seconds for Alg.~\ref{par-ctr-abs.master.alg} (i.e., the master execution time, since it wraps the overall parallel computation), 
column \textbf{CT} shows the communication time (including I/O time),
column \textbf{IO} shows the I/O time only,
column \textbf{Speedup} reports the speedup and
column \textbf{Efficiency} reports the scaling efficiency.
Finally, column \textbf{CPU K} 
shows the execution time in seconds 
for the control software generation (i.e., the remaining computation of \qks, after the control abstraction generation).

\input{pendulum-table.tex}

\subsection{Experiments Discussion}
\label{sect:expresdiscussion}

From Figs.~\ref{fig:pendulum_ctrabs_speedup.eps} and~\ref{fig:buck_ctrabs_speedup.eps} we note that the speedup is almost linear, with a $\frac{2}{3}$ slope. 
From Figs.~\ref{fig:pendulum_ctrabs_efficiency.eps} and~\ref{fig:buck_ctrabs_efficiency.eps} we note that scaling efficiency remains high when increasing the number of processors $p$. For example, for $b=22$ bits, our approach efficiency is in a range from 75\% (10 processors) to 65\% (60 processors). In any case, efficiency is always above 65\%.

\begin{figure}
\begin{tabular}{cc}
\begin{minipage}{0.45\textwidth}
\hspace*{-3mm}
	\includegraphics[scale=0.2]{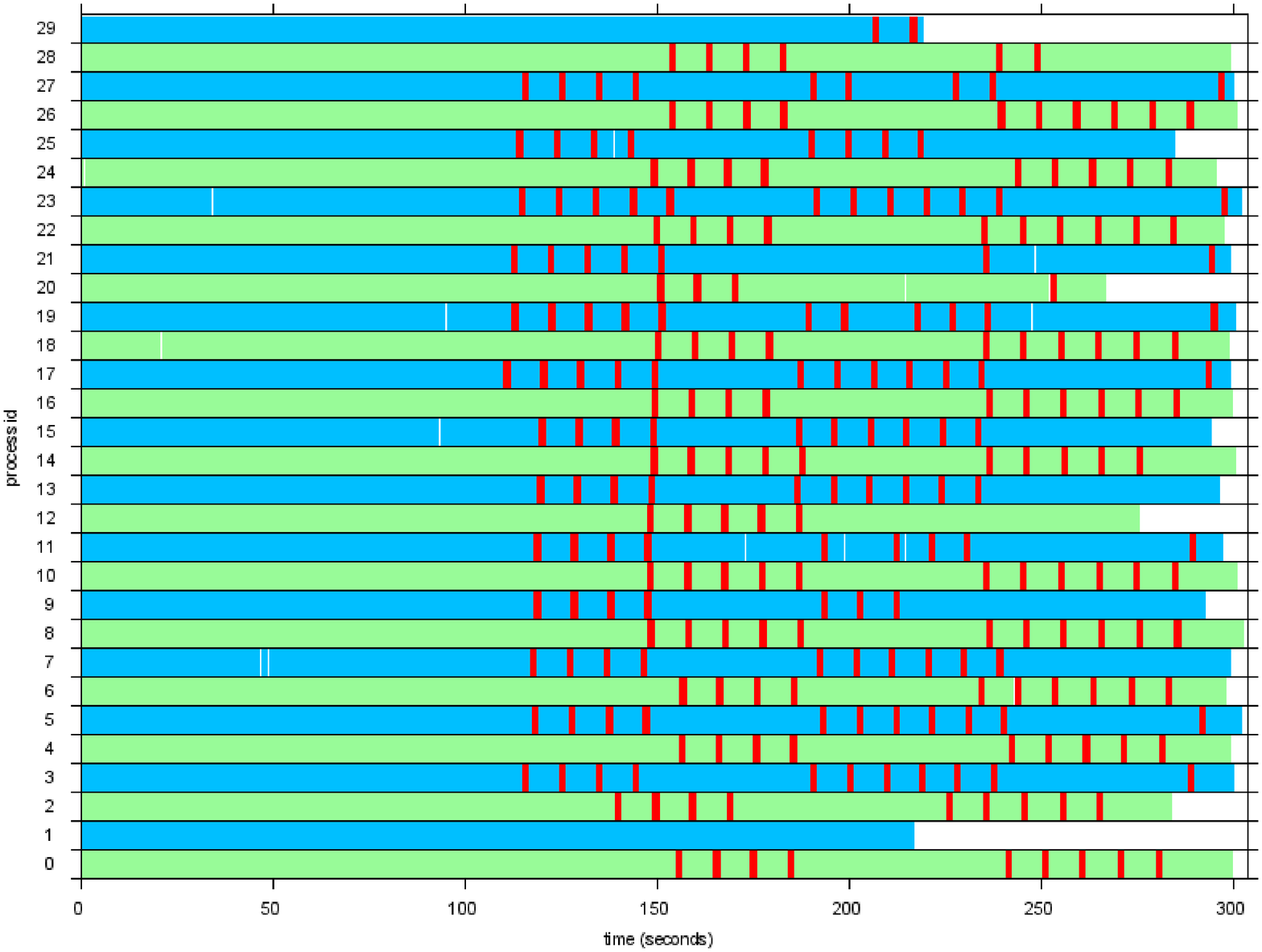}
	\caption{Details about pendulum computation time (30 nodes, 9 bits).}
	\label{fig:plot.30.clock.eps}
\end{minipage}
&
\begin{minipage}{0.45\textwidth}
\hspace*{-3mm}
	\includegraphics[scale=0.2]{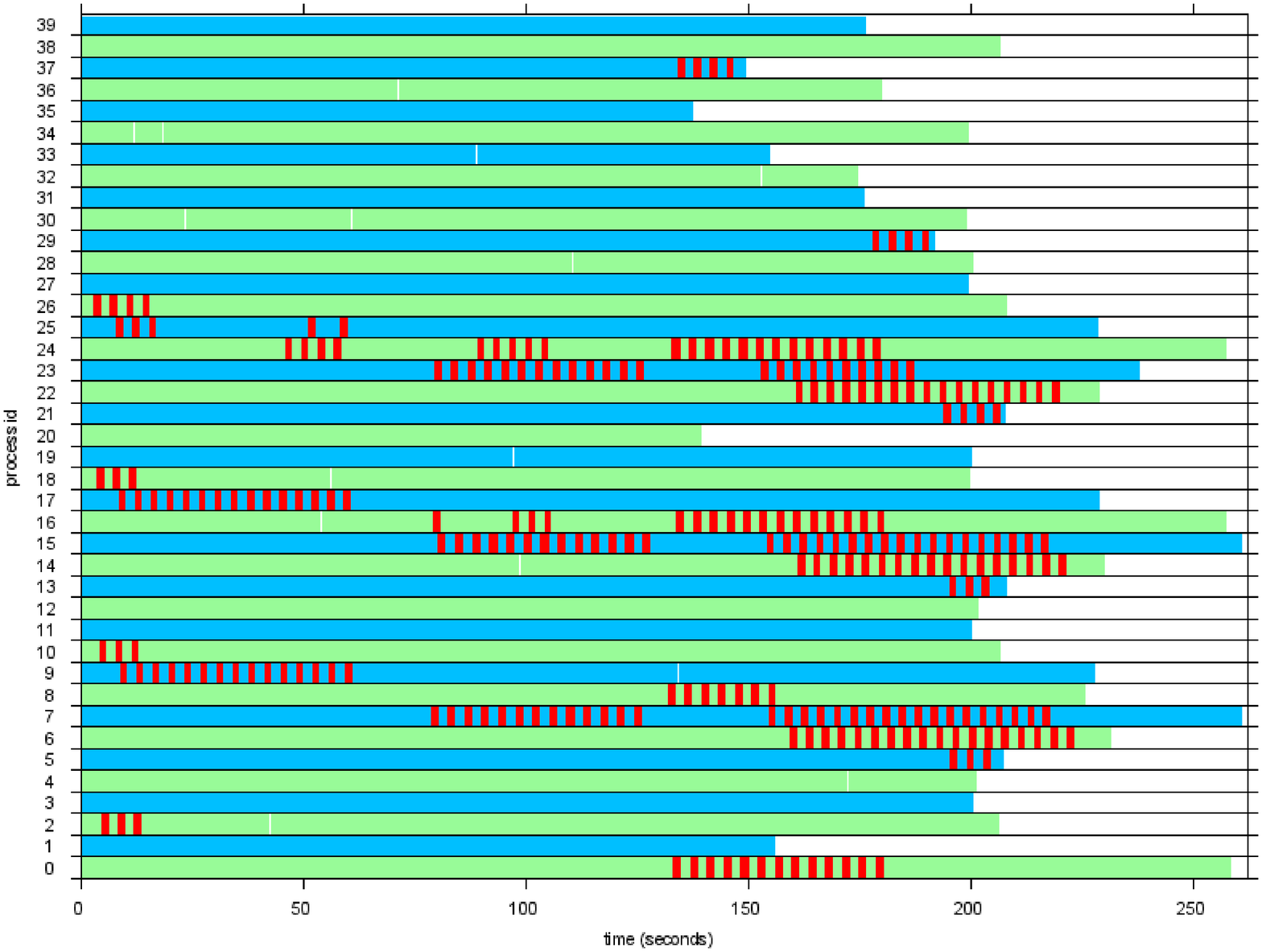}
	\caption{Details about pendulum computation time (40 nodes, 9 bits).}
	\label{fig:plot.40.clock.eps}
\end{minipage}
\end{tabular}
\end{figure}

Figs.~\ref{fig:pendulum_comm.eps} and~\ref{fig:buck_comm.eps} show that communication time almost always decreases when $p$ increases. This is motivated by the fact that, in our MPI implementation, communication among nodes takes place mostly when workers send their local control abstractions to the master via the shared filesystem. Since in our implementation this happens only after an MPI\_Barrier (i.e., the parallel computation may proceed only when all nodes have reached an MPI\_Barrier statement), the communication time also includes waiting time for workers which finishes their local computation before the other ones. Thus, if all workers need about the same time to complete the local computation, then the communication time is low. 
Note that this explains also the discontinuity when passing from 30 to 40 nodes which may be observed in the figures above. In fact, each worker has (almost) the same workload in terms of abstract states number, but some abstract states may need more computation time than others (i.e., computation time of function \fun{minCtrAbsAux} in Alg.~\ref{ctr-abs-aux.alg} may have significant variations on different abstract states). If such ``hard'' abstract states are well distributed among workers, communication time is low (with higher efficiency), otherwise it is high. Figs.~\ref{fig:plot.30.clock.eps} and~\ref{fig:plot.40.clock.eps} show such phenomenon on the inverted pendulum quantized with 18 bits, when the parallel algorithm is executed by 30 and 40 workers, respectively. In such figures, the $x$-axis represents computation time, the $y$-axis the workers, and hard abstract states are represented in red. Indeed, in Fig.~\ref{fig:plot.30.clock.eps} hard abstract states are well distributed among workers, which corresponds to a low communication time in Fig.~\ref{fig:pendulum_comm.eps} (and high speedup and efficiency in Figs.~\ref{fig:pendulum_ctrabs_speedup.eps} and~\ref{fig:pendulum_ctrabs_efficiency.eps}). On the other hand, in Fig.~\ref{fig:plot.40.clock.eps} hard abstract states are mainly distributed on only a dozen of the 40 workers (thus, about 30\% of the workers performs the most part of the real workload), which corresponds to a high communication time in Fig.~\ref{fig:pendulum_comm.eps} (and low speedup and efficiency in Figs.~\ref{fig:pendulum_ctrabs_speedup.eps} and~\ref{fig:pendulum_ctrabs_efficiency.eps}). A similar reasoning may be drawn for the I/O time.

Finally, in order to show feasibility of our approach also on DTLHSs requiring a huge computation time to generate the control abstraction, we run \pqks\ on the inverted pendulum with $b=26$. We estimate the computation time for control abstraction generation for $p = 1$ to be 25 days. On the other hand, with $p = 60$, we are able to compute the control abstraction generation in only 16 hours.

\input{buck-table.tex}

%% file: pendulum-table.tex
\begin{sidewaystable}
\centering
\small
\caption{Experimental Results for inverted pendulum.}\label{tab:invpend}
\begin{tabular}{|c|c|cccccc|c|}
\hline
\multicolumn{1}{|c|}{} & \multicolumn{1}{c|}{QKS} & \multicolumn{6}{c|}{PQKS} & \multicolumn{1}{c|}{}\\
\hline
$b$ & CPU Ctrabs & $p$ & CPU Ctrabs & CT & IO & Speedup & Efficiency & CPU K\\
\hline
18 & 8.958e+03 & 50 & 2.064e+02 & 7.696e+02 & 1.540e+01 & 43.399 & 86.798 & 2.970e+01\\
18 & 8.958e+03 & 60 & 1.763e+02 & 6.825e+02 & 1.790e+01 & 50.809 & 84.681 & 2.970e+01\\
20 & 3.108e+04 & 50 & 8.527e+02 & 3.112e+03 & 7.330e+01 & 36.450 & 72.900 & 1.131e+02\\
20 & 3.108e+04 & 60 & 7.173e+02 & 2.170e+03 & 6.740e+01 & 43.331 & 72.218 & 1.131e+02\\
22 & 1.147e+05 & 50 & 3.504e+03 & 1.242e+04 & 2.840e+02 & 32.742 & 65.485 & 1.131e+03\\
22 & 1.147e+05 & 60 & 2.938e+03 & 6.762e+03 & 2.842e+02 & 39.050 & 65.084 & 1.131e+03\\
\hline
\end{tabular}
\end{sidewaystable}

%% file: buck-table.tex
\begin{sidewaystable}
\centering
\small
\caption{Experimental Results for multi-input buck DC-DC converter.}\label{tab:buck}
\begin{tabular}{|c|c|cccccc|c|}
\hline
\multicolumn{1}{|c|}{} & \multicolumn{1}{c|}{QKS} & \multicolumn{6}{c|}{PQKS} & \multicolumn{1}{c|}{}\\
\hline
$b$ & CPU Ctrabs & $p$ & CPU Ctrabs & CT & IO & Speedup & Efficiency & CPU K\\
\hline
18 & 1.300e+05 & 50 & 4.020e+03 & 1.582e+04 & 4.100e+01 & 32.347 & 64.694 & 7.400e+01\\
18 & 1.300e+05 & 60 & 3.363e+03 & 6.550e+03 & 4.800e+01 & 38.666 & 64.443 & 7.400e+01\\
20 & 5.231e+05 & 50 & 1.619e+04 & 6.306e+04 & 1.780e+02 & 32.307 & 64.613 & 3.780e+02\\
20 & 5.231e+05 & 60 & 1.353e+04 & 2.765e+04 & 1.910e+02 & 38.657 & 64.428 & 3.780e+02\\
\hline
\end{tabular}
\end{sidewaystable}

%% file: related-work.tex
\section{Related Work}\label{related-work.tex}


Algorithms (and tools) for the
automatic synthesis of control software under different assumptions (e.g., discrete or continuous time,
linear or non-linear systems, hybrid or discrete systems, etc.) have been widely investigated in the last decades. As an example,
see~\cite{sat-opt-ctr-hscc04,icar08,upmurphi-icaps09,KB94,MazoTabuada11,jha-emsoft11,lpw_cav97,CJLRR09} and citations thereof.
However, no one of such approaches has a parallel version of any type, our focus here. 
On the other hand, parallel algorithms have been widely investigated for formal verification (e.g., see~\cite{MPSYKG09,BDLML11,BBCR10}).

A parallel algorithm for control software synthesis has been presented
in~\cite{SS98}, where however non-hybrid systems are addressed, control is
obtained by Monte Carlo simulation and quantization is not taken into account.
Moreover, note that in literature ``parallel controller synthesis'' often refers
to synthesizing parallel controllers (e.g., see~\cite{MPK10} and~\cite{PABA94} 
and citations thereof), while here we parallelize the (offline) computation
required to synthesize a standalone controller.
Summing up, to the best of our knowledge, no previous 
parallel algorithm for control software synthesis from formal specifications 
has been published.


As discussed in Sect.~\ref{motivations.subsec}, the present paper builds mainly
upon the tool \qks\ presented in~\cite{cav10,tosem13}.
Other works about \qks\ comprise the following ones. In~\cite{MMSTicsea12} 
it is shown that expressing the input system as a linear predicate
over a set of continuous as well as discrete 
variables (as it is done in \qks) is not a limitation on the modeling power. In~\cite{AMMSTcdc12} it is shown
how non-linear systems may be modeled by using suitable linearization
techniques. The paper in~\cite{AMMSTemsoft12} addresses model based synthesis of control software by trading
system level non-functional requirements (such us optimal set-up time, ripple) 
with software non-functional requirements (its footprint, i.e. size). 
The procedure which generates the actual control software (C code) starting
from a finite states automaton of a control law is described in~\cite{icsea2011}.
In~\cite{MMSTinfocomp12} it is shown
how to automatically generate a picture illustrating control software coverage.
Finally, in~\cite{MMSTictac12} it is shown that the quantized control synthesis problem underlying \qks\ 
approach is undecidable. As a consequence, \qks\ is
based on a correct but non-complete algorithm. Namely, \qks\ output is one of the
following: i) {\sc Sol}, in which case a correct-by-construction control software is returned;
ii) {\sc NoSol}, in which case no controller
exists for the given specifications; iii) {\sc Unk}, in which case \qks\ was not
able to compute a controller (but a controller may exist).


%% file: conclu.tex
\section{Conclusions and Future Work}\label{conclu.tex}

\sloppy 

In this paper we presented a Map-Reduce style parallel algorithm (and its MPI implementation for computer clusters, \pqks)
for automatic synthesis of correct-by-construction control software for discrete
time linear hybrid systems, starting from a formal model of the controlled
system, safety and liveness requirements and number of bits for
analog-to-digital conversion. Such an algorithm significantly improves
performance of an existing standalone approach (implemented in the tool \qks),
which may require weeks or even months of computation when applied to large-sized hybrid
systems.

\fussy

Experimental results on two classical control
synthesis problems (the inverted pendulum and the multi-input buck DC/DC
converter) show that our parallel approach efficiency is above 65\%.
As an example, with 60 processors \pqks\ outputs the control
software for the 26-bits quantized inverted pendulum in about 16 hours, while
\qks\ needs about 25 days of computation.
%

Future work consists in further improving the communication among processors
by making the mapping phase aware of ``hard'' abstract states (see Sect.~\ref{sect:expresdiscussion}), 
as well as designing a parallel version
for other architectures than computer clusters, such as GPGPU architectures.